\documentclass[11pt]{article}
\usepackage[margin=1in]{geometry}
\usepackage{graphicx}
\usepackage{multirow}
\usepackage{amsmath,amssymb,amsfonts}
\usepackage{amsthm}
\usepackage{rsfso}
\usepackage[title]{appendix}
\usepackage{xcolor,color}
\usepackage{textcomp}
\usepackage{manyfoot}
\usepackage{booktabs}
\usepackage{float}

\usepackage{algorithm}
\usepackage{algorithmicx}
\usepackage{algpseudocode}
\usepackage{listings}
\usepackage{amsfonts,amsmath,amssymb}
\usepackage{graphicx}
\usepackage{subcaption}
\usepackage{float}
\usepackage{placeins}
\usepackage{array}
\usepackage{hyperref}
\usepackage{cleveref}

\newcommand{\z}{\textbf{z}}
\newcommand{\zz}{\mathbf{z}}
\newcommand{\x}{\textbf{x}}

\makeatletter
\DeclareMathAlphabet{\mathcal}{OMS}{cmsy}{m}{n}
\makeatother

\theoremstyle{thmstyleone}

\theoremstyle{thmstyletwo}

\newtheorem{remark}{Remark}

\theoremstyle{thmstylethree}

\usepackage[normalem]{ulem}

\title{Perceptually Regularized Diffusion Model for Image Super-Resolution}
\author{%
\parbox{0.95\textwidth}{\centering
Chuxiangbo Wang$^{1}$,
Pavithra Venkatachalapathy$^{2}$,
Ying Liang$^{3}$,
Min Wang$^{4}$,\\
Jing Qin$^{5, }$\thanks{Corresponding author: \texttt{jing.qin@uky.edu}},
Yifei Lou$^{6}$,
Weihong Guo$^{3}$
\\[1.2ex]
\centering
\small
$^{1}$ Department of Mathematics, University of North Carolina,
Chapel Hill, NC 27599, USA.\\[0.3ex]
$^{2}$ Department of Mathematics \& Statistics, Texas Tech University,
Lubbock, TX 79409, USA.\\[0.3ex]
$^{3}$ Department of Mathematics, Applied Mathematics and Statistics,
Case Western Reserve University, Cleveland, OH 44106, USA.\\[0.3ex]
$^{4}$ Department of Mathematics, University of Houston,
Houston, TX 77204, USA.\\[0.3ex]
$^{5}$ Department of Mathematics, University of Kentucky,
Lexington, KY 40506, USA.\\[0.3ex]
$^{6}$ Department of Mathematics and School of Data Science and Society,
University of North Carolina, Chapel Hill, NC 27599, USA.
}}

\date{}

\begin{document}
\maketitle

\begin{abstract}
Image super-resolution, which aims to reconstruct high-resolution images from their low-resolution observations, is fundamental to medical imaging, remote sensing, surveillance, microscopy, and scientific visualization. Traditional model-based methods formulate super-resolution as an inverse problem with hand-crafted regularization priors. While interpretable and theoretically grounded, they rely on fixed assumptions and require computationally intensive iterative solvers. Deep learning methods offer data-driven flexibility by learning nonlinear mappings from low- to high-resolution images, among which diffusion models have achieved particularly impressive perceptual quality. However, the standard diffusion training objective is a pixel-domain noise-prediction loss that does not explicitly enforce perceptual fidelity, which can lead to oversmoothing and loss of fine image structure. To address these limitations, we propose a perceptually regularized diffusion framework that incorporates prior knowledge through perceptual-loss-based regularization, improving training convergence and encouraging the recovery of meaningful image features. Experiments on benchmark datasets demonstrate improved perceptual quality and competitive distortion metrics, highlighting the effectiveness of regularization for diffusion-based super resolution.
\end{abstract}

\vspace{0.5em}

\section{Introduction}\label{sec1}

Single image super-resolution (SR) is a fundamental image restoration task that aims to reconstruct a high-resolution (HR) image from its low-resolution (LR) observation. By recovering missing high-frequency details and improving spatial resolution, SR plays an important role in a wide range of applications, including medical imaging, remote sensing, surveillance, microscopy, scientific visualization, and digital photography \cite{wang2020deep,chen2022real}. In these domains, enhanced spatial detail can improve visual interpretation, support downstream quantitative analysis, and assist decision-making when direct acquisition of HR images is costly, time-consuming, or physically constrained.

Despite its practical importance, image super-resolution is an inherently ill-posed inverse problem: the image degradation process removes high-frequency information through downsampling, blurring, compression, and sensor noise. Thus, a single LR image may correspond to many plausible HR images. Traditional model-based methods address this ill-posedness by explicitly formulating the degradation process and incorporating hand-crafted regularization that encodes prior assumptions on image structure, such as sparsity in a transform domain, smoothness, or
total variation \cite{rudin1992nonlinear}. These methods are interpretable and mathematically grounded, but their performance is often limited by simplified assumptions about natural images and image degradation. Moreover, many classical approaches require iterative optimization, which can be computationally expensive and difficult to scale to complex real-world settings.

Deep learning has substantially advanced single-image super-resolution by learning nonlinear mappings from LR to HR images directly from data. Convolutional neural networks, residual networks, and attention-based architectures  have demonstrated remarkable performance in recovering sharper structures and more faithful textures than conventional interpolation methods such as bicubic interpolation \cite{khaledyan2020low, liu2013directional}. Generative adversarial networks (GANs)-based methods \cite{ledig2017photo, wang2018esrgan} introduce
adversarial training to encourage recovered images to lie on the manifold of natural HR images, but are susceptible to training instability, mode collapse~\cite{goodfellow2014gan,wiatrak2019stabilizing}, and visual artifacts such as unnatural textures or hallucinated structures~\cite{wang2018esrgan, xie2023desra}. More broadly, purely data-driven models may generalize poorly under unseen degradations, limited training data, or domain shifts.

Diffusion-based image super-resolution \cite{moser2024diffusion} has recently emerged as a promising alternative to GAN-based methods. Diffusion models \cite{ho2020denoising, yang2023diffusion} generate images through a progressive denoising process that transforms random noise into structured image samples. When applied to SR, the reverse diffusion process is conditioned on the LR image to gradually reconstruct an HR image that is both visually realistic and consistent with the observed LR input \cite{saharia2022image, li2022srdiff, yue2023resshift, shang2024resdiff}. Unlike deterministic regression methods, diffusion models can represent the one-to-many nature of SR by sampling multiple plausible HR reconstructions for the same LR observation, and they synthesize photorealistic textures with fewer artifacts than GAN-based
approaches~\cite{saharia2022image, wang2018esrgan}. However, the standard diffusion training objective is a pixel-domain
noise-prediction loss that does not explicitly enforce perceptual fidelity, which can lead to oversmoothing and loss of fine image
structure. Furthermore, the strong generative capability of diffusion models can produce visually plausible but physically or semantically
inaccurate details under severe degradation or out-of-distribution conditions~\cite{moser2024diffusion}, a hallucination risk that is
especially problematic in fidelity-critical applications such as medical imaging and microscopy.

Although diffusion-based SR methods have shown strong generative capability,  key challenges remain. First, the noise-prediction objective does not explicitly enforce perceptual fidelity, which can lead to oversmoothing and loss of fine image structure despite visually plausible outputs. Second, diffusion models typically require many training iterations to learn a stable denoising trajectory, and convergence to perceptually faithful
reconstructions can be slow without explicit perceptual supervision. Third, the strong generative capability can produce physically or
semantically inaccurate details under severe degradation or out-of-distribution conditions~\cite{moser2024diffusion}, a hallucination risk that is particularly problematic in fidelity-critical applications such as medical imaging and microscopy. Fourth, diffusion models may be sensitive to training data distribution, degradation mismatch, and noise levels, though this limitation is not the focus of the present work. These
challenges motivate the development of perceptually regularized diffusion frameworks that address fidelity and convergence while preserving the generative strength of the diffusion process.

In this work, we propose a perceptually regularized diffusion framework for single-image super-resolution, built on the Image
Super-Resolution via Iterative Refinement (SR3) backbone~\cite{saharia2022image}. Specifically, we augment the SR3 training
objective with a perceptual regularization term that penalizes feature-space discrepancy between the reconstructed and ground-truth
HR images, encouraging the recovery of perceptually meaningful structures without modifying the diffusion architecture or inference
procedure. As a concrete instantiation, we adopt a VGG perceptual loss~\cite{Simonyan2015VGG} to extract feature representations that capture perceptually meaningful image structures. Specifically, we use a pretrained VGG-19 network as a fixed feature extractor, which has been widely used in image restoration tasks~\cite{johnson2016perceptual,ledig2017photo, wang2018esrgan}. While the inference cost of the diffusion process is unchanged, the proposed regularization substantially reduces the number of training epochs required to reach competitive performance, offering a practical advantage in training efficiency. We further provide a gradient-based interpretation showing that the perceptual term introduces anisotropic curvature and SNR-dependent gradient scaling, offering a theoretical interpretation consistent with the accelerated early-stage convergence observed empirically. Experiments on benchmark datasets demonstrate improved perceptual quality and competitive distortion metrics, confirming the effectiveness of perceptual regularization for diffusion-based super-resolution.

The rest of the paper is organized as follows. In Section~\ref{sec:sec2}, a brief overview of diffusion methods for SR is provided. Section~\ref{sec:method} presents the proposed perceptually regularized diffusion model, including the VGG-based perceptual regularization term and a gradient-based interpretation of the accelerated convergence observed in training. In Section~\ref{sec:numerical}, we provide a variety of numerical experiments to showcase the performance of the proposed method on various datasets, including grayscale and color images.

\section{Related Work}\label{sec:sec2}
\subsection{Denoising Diffusion Probabilistic Models}\label{sec:ddpm}

Diffusion models (DMs) \cite{sohl2015deep,ho2020denoising,song2021scorebased} have emerged as a powerful class of deep generative models \cite{goodfellow2014gan,kingma2014vae,rezende2015flows} for image restoration, formulating these tasks as the reversal of a gradual noise-corruption process. Specifically, the forward diffusion process defines a joint distribution over the full trajectory $(\z_0, \z_1, \cdots, \z_T)$, i.e.,
\begin{equation}\label{eq:forward-diffusion}
  q(\z_0, \z_1, \cdots, \z_T)=q(\z_0)\prod_{t=1}^T q(\z_t|\z_{t-1}).
\end{equation}
Here, $q(\z_0)$ (usually unknown) is the data distribution of a clean image $\z_0$,  and each Markov transition kernel $q(\z_t|\z_{t-1})$  progressively corrupts $\z_0$ by adding Gaussian noise,
\begin{equation}\label{eq:true-forward-dist}
q(\z_t|\z_{t-1})
=
\mathcal{N}\!\left(
\z_t \mid \sqrt{1-\alpha_t}\z_{t-1},\alpha_t I
\right),
\end{equation}
where \(\alpha_t\in(0,1)\) is a fixed, monotonically decreasing noise schedule and $I$ denotes the identity matrix. The scaling factor $\sqrt{1-\alpha_t}$ is chosen to satisfy the variance-preserving constraint \cite{ho2020denoising,song2021scorebased}: since the signal and noise contributions have squared coefficients $(1-\alpha_t)$ and $\alpha_t$ that sum to unity, $\z_t$ retains unit variance at every timestep provided $\z_0$ is unit-variance.
By marginalizing over intermediate steps, the noisy sample at any arbitrary time step \(t\) can be written in closed form as
\[
q(\z_t|\z_0)
=
\mathcal{N}\!\left(
\z_t \mid \sqrt{\gamma_t}\z_0,(1-\gamma_t)I
\right),
\qquad
\gamma_t=\prod_{i=1}^{t}(1-\alpha_i),
\]
or equivalently via the reparameterization trick,
\begin{equation}\label{eq:reparam-z}
\z_t=\sqrt{\gamma_t}\z_0+\sqrt{1-\gamma_t}\epsilon,
\qquad
\epsilon\sim\mathcal{N}(\mathbf{0},I).
\end{equation}
For sufficiently large \(T\), the accumulated noise destroys most image structure and the marginal distribution of $\z_T$ approaches a standard Gaussian, i.e., \(\z_T\sim \mathcal{N}(\mathbf 0,I)\).

In principle, given the true reverse $q(\z_{t-1}|\z_t)$, one could recover a clean sample by initializing $\z_T\sim \mathcal N(\mathbf 0, I)$ and iteratively sampling $\z_{t-1}\sim q(\z_{t-1}|\z_t)$ for $t=T, T-1, \cdots, 1,$ yielding $\z_0\sim q(\z_0)$.
However, the true reverse $q(\z_{t-1}|\z_t)$ is intractable, but conditioning additionally on $\z_0$ yields an explicit Gaussian posterior via Bayes' rule:
$$q(\z_{t-1}|\z_t,\z_0) = \mathcal{N}\!\left(\z_{t-1}\mid\tilde{\mu}_t(\z_t,\z_0),\,\tilde{\beta}_t I \right),$$
where the posterior variance $\tilde{\beta}_t = \frac{\alpha_t(1-\gamma_{t-1})}{1-\gamma_t}$ is fully determined by the noise schedule, and the posterior mean is
\begin{equation}\label{eq:posterior-mean}
\tilde{\mu}_t(\z_t,\z_0) = \frac{1}{\sqrt{1-\alpha_t}}\!\left(\z_t - \frac{\alpha_t}{\sqrt{1-\gamma_t}}\,\boldsymbol{\epsilon}\right),
\end{equation}
with $\boldsymbol{\epsilon}$ being the noise used to construct $\z_t$ via the reparameterization \eqref{eq:reparam-z}.

Denoising Diffusion Probabilistic Model (DDPM) \cite{ho2020denoising} approximates the true reverse with a learned Gaussian reverse process,
\begin{equation}\label{eq:learned-reverse}
p_\theta(\z_0, \z_1, \cdots, \z_T) = p(\z_T)\prod_{t=1}^T p_\theta(\z_{t-1}|\z_t).
\end{equation}
Here $p(\z_T)=\mathcal N(\mathbf 0,I)$, and each reverse transition is parameterized as
\begin{equation}\label{eq:DDPM-Backwards}
    p_\theta(\z_{t-1}|\z_t) = \mathcal N\big(\z_{t-1}\mid\mu_\theta(\z_t,\gamma_t), \Sigma_\theta(\z_t,\gamma_t)\big),
\end{equation}
where $\mu_{\theta}(\z_t,\gamma_t)$ is predicted by a neural network with learnable parameters $\theta$ and $\Sigma_\theta$ is a fixed, time-dependent scalar multiple of $I.$
The reverse process is then optimized by maximum likelihood. Ideally, the training is guided by minimizing the expected negative log-likelihood of the clean data under the learned model,
\begin{equation}\label{eq:NLL}
\mathbb{E}_{q(\z_0)}\!\left[-\log p_\theta(\z_0)\right],
\qquad
p_\theta(\z_0)=\int p_\theta(\z_0,\z_1,\cdots,\z_T)\,\mathrm{d}\z_1\cdots \mathrm{d}\z_T .
\end{equation}
This marginalization is intractable, so we bound \eqref{eq:NLL} from above.
Introducing the forward process as an importance distribution and applying Jensen's
inequality to the concave function $\log(\cdot)$ gives
\begin{equation}\label{eq:Jensen}
\begin{split}
-\log p_\theta(\z_0)
&= -\log
\mathbb{E}_{q(\z_1,\cdots,\z_T|\z_0)}
\left[\frac{p_\theta(\z_0,\z_1,\cdots,\z_T)}{q(\z_1,\cdots,\z_T|\z_0)}\right]\\
&\leq
\mathbb{E}_{q(\z_1,\cdots,\z_T|\z_0)}
\left[-\log\frac{p_\theta(\z_0,\z_1,\cdots,\z_T)}{q(\z_1,\cdots,\z_T|\z_0)}\right].
\end{split}
\end{equation}
Substituting the factorizations \eqref{eq:forward-diffusion} and
\eqref{eq:learned-reverse} into the KL divergence, then taking the
expectation over $\z_0\sim q(\z_0)$, yields the variational bound
\begin{equation}\label{eq:VB}
\mathbb{E}_{q(\z_0)}\!\left[-\log p_\theta(\z_0)\right]
\;\leq\;
\underbrace{\mathbb{E}_{q(\z_0,\cdots,\z_T)}
\left[
-\log p(\z_T)
-\sum_{t=1}^{T}\log\frac{p_\theta(\z_{t-1}|\z_t)}{q(\z_t|\z_{t-1})}
\right]}_{=:\;\mathcal{L}_{\mathrm{VB}}(\theta)} ,
\end{equation}
so minimizing $\mathcal{L}_{\mathrm{VB}}$ over $\theta$ maximizes
a lower bound on the log-likelihood of clean images $\z_0$ under
the learned reverse process.
We now decompose $\mathcal{L}_{\mathrm{VB}}$ into per-timestep terms. The
decomposition is expressed through the Kullback--Leibler (KL) divergence
\[
\mathrm{KL}(q\,\|\,p)=\mathbb{E}_{q}\!\left[\log\frac{q(\mathbf{u})}{p(\mathbf{u})}\right],
\]
a nonnegative measure of the discrepancy between two distributions $q$ and $p$ that
vanishes if and only if $q=p$. Separating
the $t=1$ summand and applying Bayes' rule to the remaining ones, the Markov
property $q(\z_t|\z_{t-1})=q(\z_t|\z_{t-1},\z_0)$ gives, for $t\geq 2$,
\begin{equation}\label{eq:bayes}
\log\frac{p_{\theta}(\z_{t-1}|\z_t)}{q(\z_t|\z_{t-1})}
= \log\frac{p_{\theta}(\z_{t-1}|\z_t)}{q(\z_{t-1}|\z_t,\z_0)}
+ \log\frac{q(\z_{t-1}|\z_0)}{q(\z_t|\z_0)}.
\end{equation}
The second logarithm telescopes over $t=2,\dots,T$ to
$\log\!\big(q(\z_1|\z_0)/q(\z_T|\z_0)\big)$, whose numerator cancels the
$-\log q(\z_1|\z_0)$ contributed by the $t=1$ summand. Collecting the remaining
pieces,
\begin{equation}\label{eq:VB-decomp}
\begin{split}
\mathcal{L}_{\mathrm{VB}}(\theta)
&=\underbrace{\mathbb{E}_{q}\,\mathrm{KL}\!\left(q(\z_T|\z_0)\,\|\,p(\z_T)\right)}_{\text{prior matching}}
+\sum_{t=2}^{T}
\underbrace{\mathbb{E}_{q}\,\mathrm{KL}\!\left(q(\z_{t-1}|\z_t,\z_0)\,\|\,p_\theta(\z_{t-1}|\z_t)\right)}_{\text{denoising matching}}\\
&\qquad\quad -\underbrace{\mathbb{E}_{q}\!\left[\log p_\theta(\z_0|\z_1)\right]}_{\text{reconstruction}}.
\end{split}
\end{equation}
The prior matching term compares the endpoint of the forward process with the
Gaussian prior $p(\z_T)$ and contains no learnable parameters, so it is discarded.
The reconstruction term is absorbed into the denoising matching terms by letting the
sum run from $t=1$, following \cite{ho2020denoising}. Training is therefore governed
by the denoising matching terms,
which involve $\mathrm{KL}(q(\z_{t-1}|\z_t,\z_0)\,\|\, p_\theta(\z_{t-1}|\z_t))$ at each timestep. Since both distributions are Gaussian with the same fixed variance $\tilde{\beta}_t I$,
the KL divergence at each step reduces up to a constant factor to the squared Euclidean distance between their mean vectors:
$$\mathrm{KL}\bigg(q(\z_{t-1}|\z_t,\z_0)\,\|\,p_\theta(\z_{t-1}|\z_t)\bigg) \propto \|\tilde{\mu}_t(\z_t,\z_0) - \mu_\theta(\z_t,\gamma_t)\|_2^2.$$
Here $\|\cdot\|_2$ denotes the $\ell_2$-norm of a vector and $\|\mathbf{x}-\mathbf{y}\|_2$ denotes the Euclidean distance between the vectors $\mathbf{x}$ and $\mathbf{y}$. We parameterize $\mu_\theta$ in the same functional form as the posterior mean $\tilde{\mu}_t$ in \eqref{eq:posterior-mean}, with $\boldsymbol{\epsilon}$ replaced by the network prediction $\epsilon_\theta$, thus leading to
\begin{equation}
   \mu_\theta(\z_t,\gamma_t) = \frac{1}{\sqrt{1-\alpha_t}}\!\left(\z_t - \frac{\alpha_t}{\sqrt{1-\gamma_t}}\,\epsilon_\theta(\z_t,\gamma_t)\right).
   \label{eq:DDPM-mean}
\end{equation}
Substituting  $\tilde{\mu}_t$ and $\epsilon_\theta$ into the squared difference of means gives
 $$\|\tilde{\mu}_t(\z_t,\z_0) - \mu_\theta(\z_t,\gamma_t)\|_2^2 = \frac{\alpha_t^2}{(1-\alpha_t)(1-\gamma_t)}\|\boldsymbol{\epsilon} - \epsilon_\theta(\z_t,\gamma_t)\|_2^2.$$
Let $\lambda(t)$ absorb the scalar prefactor
$\frac{\alpha_t^2}{(1-\alpha_t)(1-\gamma_t)}$, and let
$\mathcal{U}(1,T)$ denote the uniform distribution over timesteps
$\{1,2,\ldots,T\}$. Taking the expectation of the per-step loss
over $t\sim\mathcal{U}(1,T)$, $\z_0\sim q(\z_0)$, and
$\boldsymbol{\epsilon}\sim\mathcal{N}(\mathbf{0},\mathbf{I})$
yields the noise-prediction objective:
\begin{equation}\label{eq:DDPM_loss_0}
\mathbb{E}_{\substack{
t\sim \mathcal{U}(1,T)\\
\z_0\sim q(\z_0)\\
\boldsymbol{\epsilon}\sim \mathcal{N}(\mathbf{0},\mathbf{I})
}}
\left[
\lambda(t)
\left\|
\boldsymbol{\epsilon}
-
\epsilon_{\theta}(\z_t,\gamma_t)
\right\|_2^2
\right],
\end{equation}
where $\epsilon_\theta(\z_t,\gamma_t)$ predicts the Gaussian noise used to construct $\z_t$. Note that $\z_{t}$ depends implicitly on $\z_0$ and
$\boldsymbol{\epsilon}$ through $\z_t$ via \eqref{eq:reparam-z},
which is why the expectation in \eqref{eq:DDPM_loss_0} is taken
over all three random variables $t$, $\z_0$, and $\boldsymbol{\epsilon}$. In practice, following \cite{ho2020denoising}, the weighting is set to $\lambda(t)=1$, leading to the simplified objective
\begin{equation}\label{eq:DDPM_loss}
\mathbb{E}_{\substack{
t\sim \mathcal{U}(1,T)\\
\z_0\sim q(\z_0)\\
\boldsymbol{\epsilon}\sim \mathcal{N}(\mathbf{0},\mathbf{I})
}}
\left[
\left\|
\boldsymbol{\epsilon}
-
\epsilon_{\theta}(\z_t,\gamma_t)
\right\|_2^2
\right].
\end{equation}

After minimizing \eqref{eq:DDPM_loss} with respect to $\theta$, the trained noise predictor $\epsilon_\theta$ is used to iteratively sample the reverse process from $\z_T\sim\mathcal{N}(\mathbf{0},I)$ to $\z_0$ by progressively denoising at each timestep: compute mean $\mu_{\theta}$ by formula (\ref{eq:DDPM-mean}),  and then sample with fixed $\Sigma_\theta$ based on (\ref{eq:DDPM-Backwards})
for $t=T,T-1,\ldots,1$, progressively denoising $\z_T$ to obtain a clean sample $\z_0$.
DDPM sampling is therefore an iterative denoising procedure: at each step, the model estimates the noise component in $\z_t$, removes part of it, and injects calibrated Gaussian uncertainty through $\Sigma_\theta$.

\subsection{Diffusion Models for Super-Resolution}\label{sec:sr3}

SR3 \cite{saharia2022image} extends DDPM to the conditional SR setting by incorporating a LR observation $\x$ to guide the reverse process. Without such conditioning, the learned reverse process $p_{\theta}(\z_{t-1}|\z_t)$ generates arbitrary natural images consistent with the learned data distribution, but with no guarantee of fidelity to any specific LR input. SR3 addresses this by conditioning each reverse transition on $\x:$
\begin{equation}\label{eq:SR3reverse}
p_{\theta}(\z_{t-1}|\z_t,\x)
=
\mathcal{N}\!\left(
\z_{t-1}
\mid
\mu_{\theta}(\x,\z_t,\gamma_t),\,
\Sigma_{\theta}(\x,\z_t,\gamma_t)
\right),
\end{equation}
where $\mu_{\theta}(\x,\z_t,\gamma_t)$ and $\Sigma_{\theta}(\x,\z_t,\gamma_t)$ denote the mean and covariance of the learned reverse transition, extending the DDPM parameterization \eqref{eq:DDPM-Backwards}  to incorporate the LR conditioning image $\x.$

In practice, $\x$ is first upsampled once to the target resolution by bicubic interpolation; this upsampled image is then fixed and concatenated with the noisy image $\z_t$ along the channel dimension at every denoising step, providing $\x$ as a persistent conditioning input throughout the reverse process. The noise predictor therefore takes the form $\epsilon_\theta(\x,\z_t,\gamma_t)$, and the role of the two inputs is complementary: $\x$ provides low-frequency structural information that constrains the global content of the output, while $\z_t$ carries the stochastic high-frequency component that the reverse process progressively refines into realistic HR details.   The DDPM objective \eqref{eq:DDPM_loss} thus extends naturally to the conditional setting
\begin{equation}\label{eqn:SR3}
\mathcal{L}_{\mathrm{diff}}(\epsilon_{\theta})
:=
\mathbb{E}_{\substack{
t\sim \mathcal{U}(1,T)\\
\z_0\sim q(\z_0)\\
\boldsymbol{\epsilon}\sim \mathcal{N}(\mathbf{0},\mathbf{I})
}}
\left[
\left\|
\boldsymbol{\epsilon}
-
\epsilon_{\theta}(\x,\z_t,\gamma_t)
\right\|_2^2
\right],
\end{equation}
where the subscript ``diff'' stems from diffusion. For later use, we define $\z_{\theta,t}$ as the reconstructed image at timestep $t$ induced by the noise predictor.

After minimizing \eqref{eqn:SR3} with respect to $\theta$, the trained noise predictor $\epsilon_\theta$ is used to iteratively sample the reverse process following \eqref{eq:SR3reverse} from $\z_T\sim\mathcal{N}(\mathbf{0},I)$ to $\z_0$ by progressively denoising at each timestep. SR3 forms the foundation of our proposed method described in Section \ref{sec:method}. Other notable diffusion-based SR methods include SRDiff \cite{li2022srdiff}, which trains the diffusion model to generate the residual between the HR and upsampled LR images rather than the full HR image, using a residual-in-residual dense block \cite{wang2018esrgan} as the conditioning encoder. Subsequent works such as ResShift~\cite{yue2023resshift}
and ResDiff~\cite{shang2024resdiff} further explore residual diffusion and efficient sampling strategies to improve reconstruction quality and reduce computational cost.

\section{Proposed Regularized Diffusion Methods}\label{sec:method}
The standard diffusion objective $\mathcal{L}_{\mathrm{diff}}$ in \eqref{eqn:SR3} is formulated as a noise-prediction loss in the pixel domain, which encourages the model to reverse the forward noise-corruption process but does not explicitly enforce the preservation of perceptually meaningful
image features. As a result, reconstructed images may exhibit oversmoothing, loss of semantic structures, or visually inconsistent
textures despite achieving favorable pixel-wise error metrics~\cite{ledig2017photo, wang2018esrgan}. To overcome this limitation, we incorporate a perceptual regularization term into the diffusion training objective, guiding the reconstruction toward perceptually faithful HR images by measuring discrepancy in a learned feature space rather than the pixel domain. Our proposed method augments the SR3 framework with such a term, leading to the total training objective
\begin{equation}\label{eq:loss_total}
\mathcal{L}_{\mathrm{total}}
=
\mathcal{L}_{\mathrm{diff}}
+
\mathcal{L}_{\mathrm{VGG}},
\end{equation}
where $\mathcal{L}_{\mathrm{diff}}$ is the conditional diffusion loss defined in~\eqref{eqn:SR3} and $\mathcal{L}_{\mathrm{VGG}}$ is the perceptual regularization term defined in Section~\ref{sec:vgg}. An overview of the proposed training framework is provided in Figure~\ref{fig:pipeline}.

\subsection{VGG Perceptual Regularization}\label{sec:vgg}
Unlike conventional pixel-wise losses, perceptual losses measure the discrepancy between images in a feature space learned by a deep neural network. These feature representations capture higher-level image characteristics, such as edges, textures, shapes, and semantic content, which pixel-wise metrics fail to reflect. Perceptual regularization therefore encourages reconstructed images to be visually closer to the ground-truth HR image, improving texture fidelity and visual quality while preserving the generative capability of the diffusion process~\cite{johnson2016perceptual, ledig2017photo, wang2018esrgan}.

The Visual Geometry Group (VGG) network~\cite{Simonyan2015VGG} is one of the most widely used perceptual feature extractors in image
restoration. Its deep hierarchy of convolutional layers progressively encodes low-level image details and high-level semantic information,
and pretrained VGG networks have been extensively adopted as fixed feature extractors for perceptual image reconstruction
tasks~\cite{johnson2016perceptual, ledig2017photo, wang2018esrgan}. In this work, we use a pretrained VGG-19 network as a fixed perceptual
feature extractor; its parameters remain frozen throughout training, introducing no additional trainable variables into the diffusion model.

\begin{figure}[!htbp]
    \centering
    \includegraphics[width=\linewidth]{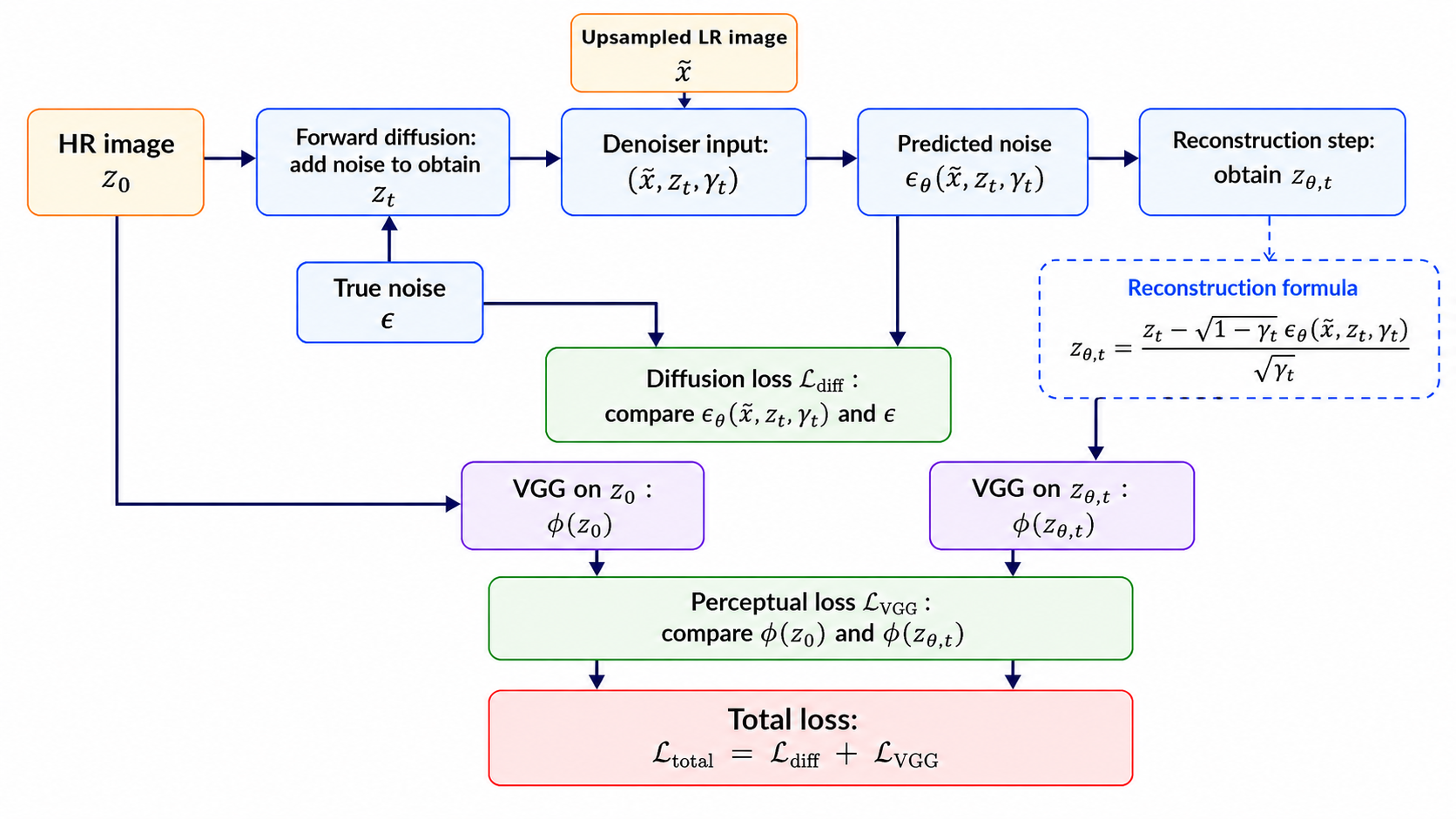}
    \caption{Overview of the proposed diffusion-based super-resolution framework with perceptual learning.
    }
    \label{fig:pipeline}
\end{figure}

Let $\phi(\cdot)$ denote the feature mapping defined by a pretrained VGG network, and $\z_0$ the corresponding ground-truth HR image. To define the perceptual loss, by rearranging the forward noising relation \eqref{eq:reparam-z}, the predicted noise $\epsilon_\theta(\x,\z_t,\gamma_t)$ gives an estimate of the HR image at each $t$:
\begin{equation}\label{eq:z-recon}
\z_{\theta,t}
:=
\frac{\z_t - \sqrt{1-\gamma_t}\,\epsilon_\theta(\x,\z_t,\gamma_t)}{\sqrt{\gamma_t}}.
\end{equation}
The resulting training pipeline is illustrated in Figure~\ref{fig:pipeline}. Given a clean HR image $\z_0$, the forward diffusion process adds noise $\boldsymbol{\epsilon}$ to produce the noisy image $\z_t$. The denoiser predicts $\epsilon_\theta$ from the upsampled LR condition and the noisy image $\z_t$. The predicted noise is used both to compute the diffusion loss $\mathcal{L}_{\mathrm{diff}}$ and, through~\eqref{eq:z-recon}, to obtain the reconstructed HR image $\z_{\theta,t}$. The perceptual loss $\mathcal{L}_{\mathrm{VGG}}$ then measures the discrepancy between the VGG feature representations of $\z_{\theta,t}$ and the ground-truth image $\z_0$.

We consider two forms of perceptual regularization term $\mathcal{L}_{\mathrm{VGG}}$, based on the $\ell_1$ and squared $\ell_2$ norm of the difference between VGG feature representations, respectively. We denote these two variants by $\mathcal{L}_{\mathrm{VGG}}^{(1)}$ and $\mathcal{L}_{\mathrm{VGG}}^{(2)}$. The $L^1$ variant is
\begin{equation}\label{eq:VGG_Loss_L1}
\mathcal{L}_{\mathrm{VGG}}^{(1)}(\epsilon_\theta)
:=
\omega_{\mathrm{VGG}}
\,\mathbb{E}_{\substack{
t\sim \mathcal{U}(1,T)\\
\z_0\sim q(\z_0)\\
\boldsymbol{\epsilon}\sim \mathcal{N}(\mathbf{0},\mathbf{I})
}}
\!\left[
\left\|
\phi(\z_{\theta,t})-\phi(\z_0)
\right\|_{1}
\right],
\end{equation}
where $\|\cdot\|_1$ denotes the $\ell_1$-norm and $\omega_{\mathrm{VGG}}>0$ is a weighting parameter controlling the strength of perceptual regularization. We distinguish the $L^1$ distance between functions (or distributions) from the $\ell_1$ norm of finite-dimensional vectors.
The second one uses the squared $L^2$ distance between VGG feature maps,
\begin{equation}\label{eq:VGG_Loss_L2}
\mathcal{L}_{\mathrm{VGG}}^{(2)}(\epsilon_\theta)
:=
\omega_{\mathrm{VGG}}
\,\mathbb{E}_{\substack{
t\sim \mathcal{U}(1,T)\\
\z_0\sim q(\z_0)\\
\boldsymbol{\epsilon}\sim \mathcal{N}(\mathbf{0},\mathbf{I})
}}
\!\left[
\left\|
\phi(\z_{\theta,t})-\phi(\z_0)
\right\|_{2}^{2}
\right].
\end{equation}
The $L^1$ formulation is less sensitive to large feature discrepancies, whereas the squared-$L^2$ formulation penalizes large feature discrepancies more strongly. Their empirical effects on reconstruction quality and convergence are compared in Section~\ref{sec:numerical}. In practice, the choice between the two formulations depends on the desired balance between perceptual sharpness and reconstruction stability.

The idea of augmenting diffusion model training with perceptual supervision has been explored in recent work. Berrada et al.~\cite{berrada2025boosting} introduced a latent perceptual loss defined over the internal features of the autoencoder decoder, showing that feature-space supervision
improves sharpness and realism in latent diffusion models without increasing inference cost. Our work pursues a similar motivation but
differs in three respects: we operate within the SR3 conditional diffusion framework for image super-resolution rather than latent
diffusion for general image synthesis; we use a pretrained VGG-19 as a simple, fixed external feature extractor rather than a decoder-internal
loss; and we study both $L^1$ and $L^2$ variants of the perceptual loss, providing empirical guidance for the choice of perceptual regularization in diffusion-based SR.

\subsection{A Gradient-Based Interpretation of the Perceptual Loss}\label{sec:gradient}

Evaluating the perceptual loss on the induced reconstruction endows the training signal with two properties: an anisotropic gradient geometry
and an implicit, SNR-dependent timestep emphasis. Together, these two properties offer a geometric interpretation consistent with the
accelerated convergence and improved perceptual fidelity observed empirically.

\textbf{Anisotropic gradient geometry.} Since gradient and expectation commute, the geometric structure derived below for the per-sample loss applies
directly to the full training objective~\eqref{eq:VGG_Loss_L2}. The per-sample squared $L^2$ perceptual loss takes the form
\begin{equation}\label{eq:vgg_multilayer}
\mathcal{L}(\z_{\theta,t})
:= \omega_{\mathrm{VGG}}
\bigl\|\phi(\z_{\theta,t})-\phi(\z_0)\bigr\|_2^2,
\end{equation}
whose Euclidean gradient with respect to $\z_{\theta,t}$ is
\begin{equation}\label{eq:vgg_gradient}
\nabla_{\z_{\theta,t}}\mathcal{L}(\z_{\theta,t})
= 2\omega_{\mathrm{VGG}}
J(\z_{\theta,t})^{\top}
\bigl(\phi(\z_{\theta,t})-\phi(\z_0)\bigr),
\end{equation}
with $J(\z_{\theta,t}) = \frac{\partial\phi}{\partial\z_{\theta,t}}$.
The Jacobian $J$ in~\eqref{eq:vgg_gradient} pulls the Euclidean metric on feature space back to pixel space, and yields a position-dependent positive-semidefinite pullback metric
\begin{equation}
G(\z_{\theta,t})
= 2\omega_{\mathrm{VGG}}
 J(\z_{\theta,t})^\top J(\z_{\theta,t}),
\label{eq:pullback_metric}
\end{equation}
which is the Gauss--Newton approximation of the Hessian of $\mathcal{L}(\z_{\theta,t})$. Thus, the perceptual loss induces a position-dependent and anisotropic local geometry in reconstruction space through $G(\z_{\theta,t})$. Directions that produce larger changes in the VGG feature representation, potentially corresponding to perceptually meaningful structures such as edges, local contrast, and texture boundaries, receive greater curvature, whereas directions to which the feature representation is relatively insensitive correspond to smaller singular values of $J$ and therefore contribute less strongly to the perceptual loss. As the reconstruction $\z_{\theta,t}$ evolves during training, $G(\z_{\theta,t})$ changes accordingly, yielding an adaptive local geometry that weights reconstruction directions according to their sensitivity in the VGG feature space.

Minimizing $\mathcal{L}(\z_{\theta,t})$ by (stochastic) gradient descent therefore behaves locally, like descent on a quadratic whose
curvature is concentrated along perceptually salient directions: the loss landscape itself encodes a data-driven  geometry on image
space. By contrast, the standard diffusion objective~\eqref{eqn:SR3} applies an isotropic Euclidean penalty to the noise-prediction error, without explicitly weighting directions according to their perceptual relevance. Adding the perceptual term to the training objective~\eqref{eq:loss_total} thus introduces anisotropic curvature into the composite loss precisely along structurally informative directions. This effect is visible in local structure rather than in aggregate pixel error: the zoomed comparisons at one million iterations (\Cref{fig:mri_late_reconstruction,fig:brainweb_8x_late_reconstruction,fig:Tumor_mri_late_reconstruction} for grayscale images and \Cref{fig:celeb_zoomed_stack_005} for color images) show better-separated adjacent structures, more continuous thin filaments, and sharper edge contrast, while the peak PSNR/SSIM values remain essentially unchanged. The $L^1$ variant~\eqref{eq:VGG_Loss_L1} induces a similar property in its (sub)gradient field, as the two variants differ in how feature discrepancies are weighted, and we compare them empirically in Section~\ref{sec:para_sensitivity}.

\textbf{Implicit timestep reweighting and accelerated early
convergence.}
The perceptual loss is evaluated not directly on the network output
$\epsilon_\theta$, but on the induced reconstruction $\z_{\theta,t}$
defined in~\eqref{eq:z-recon}. By the chain rule, the gradient of $\mathcal L$ in~\eqref{eq:vgg_multilayer} with respect to the network
output factorizes as
\begin{equation}
\frac{\partial \mathcal{L}}{\partial \epsilon_\theta}
  \;=\;
  \frac{\partial \mathcal{L}}{\partial \z_{\theta,t}}
  \cdot
  \frac{\partial \z_{\theta,t}}{\partial \epsilon_\theta},
  \qquad
  \frac{\partial \z_{\theta,t}}{\partial \epsilon_\theta}
  \;=\;
  -\,\frac{\sqrt{1-\gamma_t}}{\sqrt{\gamma_t}}I
  \;=\;
  -\,\frac{1}{\sqrt{\mathrm{SNR}(t)}}I,
\label{eq:snr_factor}
\end{equation}
where $I$ is the identity matrix of appropriate dimension, and $\mathrm{SNR}(t) = \gamma_t/(1-\gamma_t)$ is the signal-to-noise ratio (SNR) of the forward process.

\emph{Low noise.} When $\gamma_t$ is close to one, the scalar factor $\mathrm{SNR}(t)^{-1/2}$ is small, so the perceptual gradient propagated
to $\epsilon_\theta$ is attenuated. When the noise prediction is reasonably accurate, the induced reconstruction $\zz_{\theta,t}$ is also
close to $\zz_0$, further reducing the perceptual residual and its contribution to the training gradient.

\emph{High noise.} As $\gamma_t$ decreases, the scalar factor $\mathrm{SNR}(t)^{-1/2}$ increases, amplifying the perceptual gradient
propagated from $\zz_{\theta,t}$ to $\epsilon_\theta$. Moreover, higher-noise timesteps generally correspond to more challenging
reconstruction conditions, under which the perceptual residual may also be larger. Thus, the perceptual term tends to exert a stronger influence
at noisier timesteps, although its exact gradient magnitude also depends on the current reconstruction and the local VGG Jacobian. Because the
schedule keeps $\gamma_t$ bounded away from zero, the SNR-dependent scaling factor remains finite throughout training. The perceptual
regularizer therefore induces an \emph{implicit, timestep-dependent scaling of the perceptual gradient} that is related to SNR-based loss
reweighting, arising from the reconstruction mapping rather than from an explicitly prescribed timestep weight.

\emph{Structure of the signal.} Over the range of noise levels for which $\z_{\theta,t}$ remains a plausible image, this
gradient is not merely larger in magnitude but \emph{structured}: as
established by the metric~\eqref{eq:pullback_metric}, it guides $\zz_{\theta,t}$ toward the ground truth along directions to which the VGG features are most sensitive. Early in training, $\epsilon_\theta$ is still an inaccurate predictor of the injected noise, and the resulting denoising trajectory may therefore remain poorly structured. This is the regime in which the empirical gap is largest: across all four experimental settings the convergence curves separate most at the first evaluation checkpoint (see \Cref{fig:vgg_convergence_test,fig:brainweb_2x_convergence,fig:brainweb_8x_convergence,fig:tumor_4x_convergence}), and the corresponding reconstructions at 100K iterations (see \Cref{fig:mri_early_reconstruction,fig:mri_early_reconstruction_2x,fig:brainweb_8x_convergence_image_compare,Tumor_convergence_image_compare}) show the baseline still dominated by noise while the regularized model has already recovered coherent structure.

The following remark summarizes the gradient-level explanation for the accelerated convergence and improved perceptual fidelity observed
empirically in Section~\ref{sec:numerical}.

\begin{remark}\label{rem:snr_emphasis}
The anisotropic geometry in~\eqref{eq:pullback_metric} and the SNR-dependent scaling in~\eqref{eq:snr_factor} together show that the perceptual term introduces structured, timestep-dependent gradient modulation. This mechanism is consistent with the faster early-stage convergence and better-preserved local structures observed empirically relative to the unregularized baseline.
\end{remark}

\section{Numerical Examples}\label{sec:numerical}

We evaluate the effect of the proposed VGG-regularized SR3 variants $\mathcal{L}_{\mathrm{VGG}}^{(1)}, \mathcal{L}_{\mathrm{VGG}}^{(2)}$ and compare them with the baseline SR3 model without perceptual regularization across both grayscale medical images and RGB facial images. The numerical experiments include evaluations on the BrainWeb MRI dataset\footnote{Available at \url{https://brainweb.bic.mni.mcgill.ca}} \cite{Cocosco1997BrainWebOI,Kwan1999MRI,kwan1996extensible,Collins1998Design} at multiple scale factors, experiments on
the Brain Tumor MRI dataset\footnote{Available at \url{https://www.kaggle.com/datasets/masoudnickparvar/brain-tumor-mri-dataset}}, and a cross-dataset facial-image experiment in which the models are trained on FFHQ\footnote{Available at \url{https://github.com/nvlabs/ffhq-dataset}}  \cite{karras2019stylebased} and evaluated on CelebA-HQ\footnote{Available at \url{https://www.kaggle.com/datasets/badasstechie/celebahq-resized-256x256}} \cite{liu2015faceattributes,karras2017progressive}. An overview of all of the datasets, super-resolution tasks, and VGG configurations is provided in \Cref{tab:numerical_experiment_overview}.

\begin{table}[ht]
\centering
\renewcommand{\arraystretch}{1.4}
\begin{tabular}{l|l|l|c} \hline
\textbf{Experiment} & \textbf{Training data} & \textbf{Evaluation data} & \textbf{Scale} \\ \hline
Grayscale & BrainWeb MRI (3001) & Held-out BrainWeb MRI (100) & $2\times$, $4\times$, $8\times$ \\ \hline
Grayscale & Brain Tumor MRI (4712) & Held-out Brain Tumor MRI (100) & $4\times$ \\ \hline
Facial images & FFHQ (2585) & CelebA-HQ (100 of 500) & $4\times$ \\ \hline
\end{tabular}
\caption{Overview of the dataset description and configuration used in  the numerical experiments. Here, ``held-out" means the evaluation dataset is disjoint from the training set, although both sets are obtained from the same source. The values in parentheses indicate the number of images used in each training or evaluation set.}
\label{tab:numerical_experiment_overview}
\end{table}

All models are trained for one million iterations, optimized with Adam (learning rate $1\times10^{-4}$, batch size $2$) using a $2000$-step diffusion process with a linear beta schedule. Reconstruction quality is evaluated using peak signal-to-noise ratio (PSNR) and structural similarity index measure (SSIM). In addition to evaluating the reconstructions at the end of the one million iterations, we examine early convergence as well as quantitative and qualitative metrics.  For selected experiments, we also evaluate the stability of the results through multiple independent training runs. All grayscale MRI experiments are performed on an NVIDIA A100 GPU with target HR image size $128 \times 128$, whereas the RGB facial image experiments are performed on an NVIDIA V100 GPU with target HR size $128 \times 128$.

\subsection{Grayscale MRI Super-Resolution}\label{subsec:gray_mri}

We first present the grayscale MRI results. BrainWeb MRI is evaluated at scale factors $2\times$, $4\times$, and $8\times$, with the $4\times$ task serving as the main setting. The $2\times$ and $8\times$ tasks examine the behavior of VGG regularization under different degradation levels, while the Brain Tumor MRI $4\times$ task evaluates its performance on a different medical image dataset. The parameter sensitivity is presented separately in \Cref{sec:para_sensitivity}.

\subsubsection{BrainWeb MRI: detailed study on the \texorpdfstring{$4\times$}{4x} task} \label{BrainWeb MRI: 4x}

The BrainWeb MRI $4\times$ task is examined in detail as the primary experimental setting. Since the weight yielding the best final reconstruction quality does not necessarily provide the fastest early convergence, we report the models selected according to these two criteria separately.

The comparison of the selected VGG weights is summarized in \Cref{tab:vgg_best_l1_vs_l2_mri}. For final PSNR, the baseline SR3 model achieves a slightly better value. For SSIM, the $\mathcal{L}^{(2)}_{\mathrm{VGG}}$ model with $\omega_{\mathrm{VGG}}=1$ slightly outperforms SR3.

\begin{figure}[!htbp]
    \centering
    \begin{subfigure}{0.48\linewidth}
        \centering
        \includegraphics[width=\linewidth]{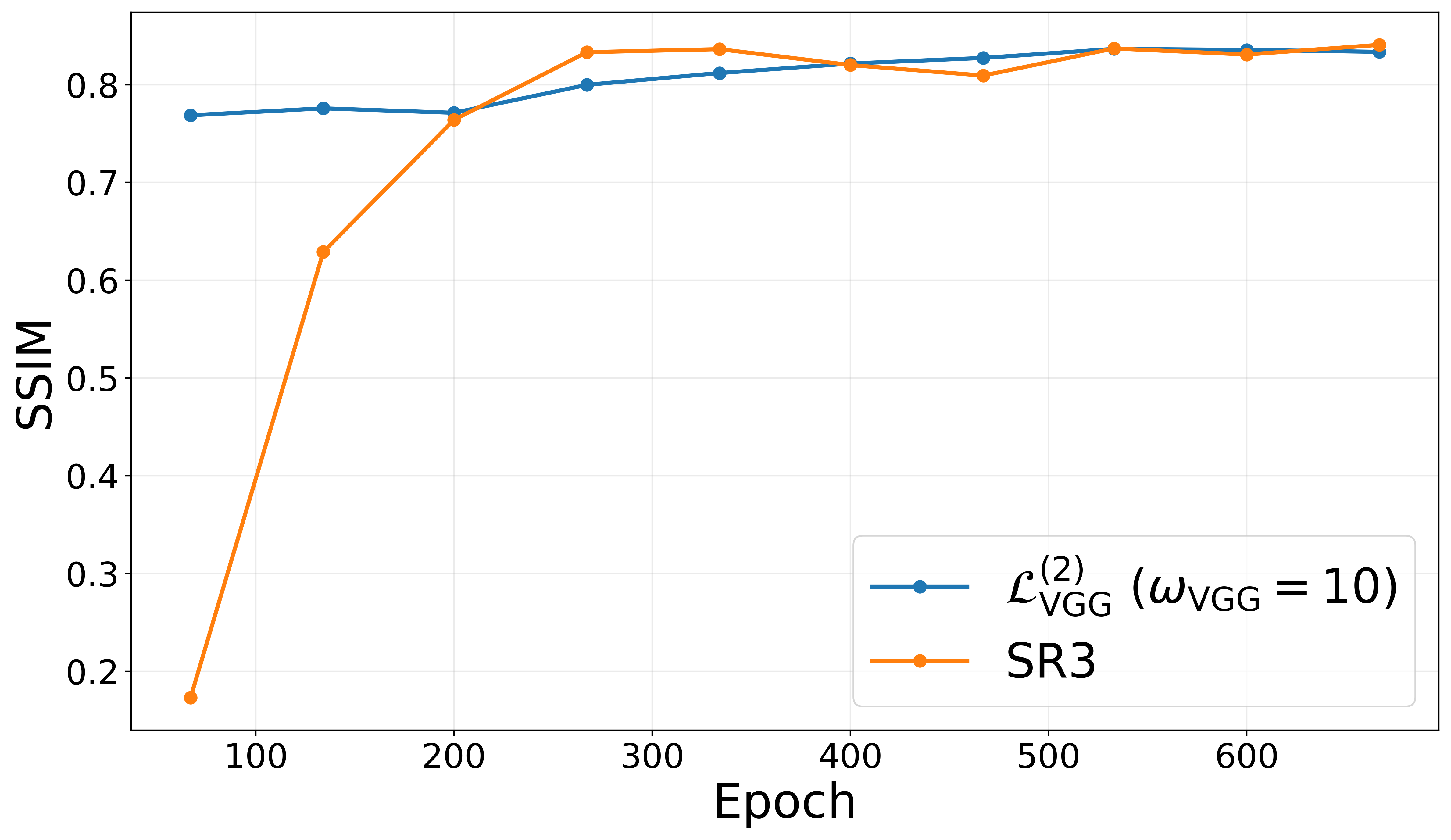}
        \caption{SSIM}
        \label{fig:vgg_ssim_convergance}
    \end{subfigure}
    \hfill
    \begin{subfigure}{0.48\linewidth}
        \centering
        \includegraphics[width=\linewidth]{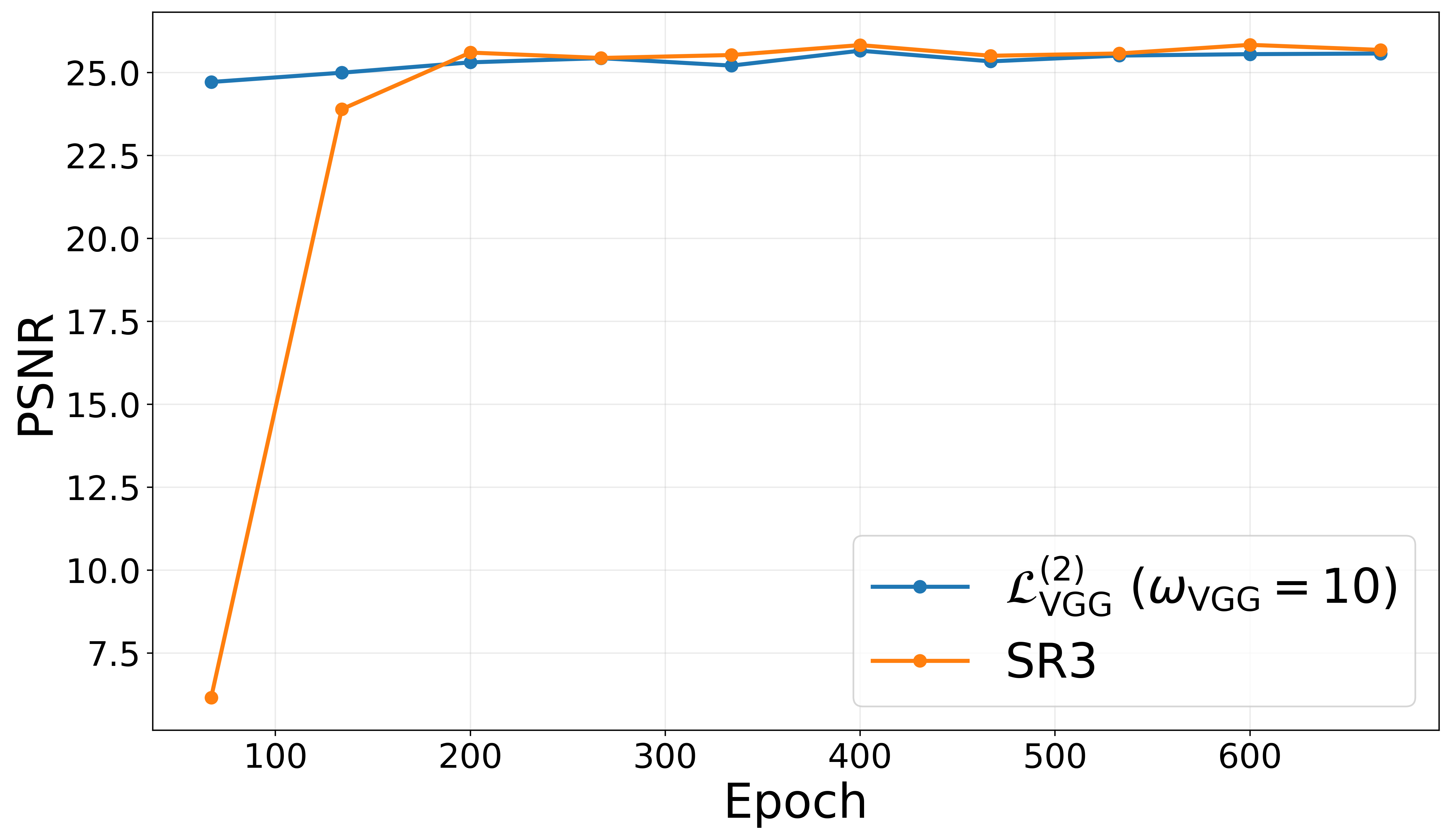}
        \caption{PSNR}
        \label{fig:vgg_psnr_converges}
    \end{subfigure}
    \caption{Convergence comparison between the proposed method ($\mathcal{L}_{\mathrm{VGG}}^{(2)}$, $\omega_{\mathrm{VGG}}=10$; blue) and the baseline SR3 model (orange).}
    \label{fig:vgg_convergence_test}
\end{figure}

\begin{table}[htbp]
    \centering
     \begin{tabular}{c|c|c|c|c}
    \hline
     Model & Selected weight (quality) & Selected weight (convergence) & Peak PSNR & Peak SSIM \\
     \hline
     $\mathcal{L}^{(2)}_{\mathrm{VGG}}$ & 1 &-- & 25.753 & \textbf{0.8436} \\
     $\mathcal{L}^{(2)}_{\mathrm{VGG}}$ & -- & 10& 25.660 & 0.8365 \\
      SR3 & -- & -- &\textbf{25.837} &0.8406  \\
     \hline
    \end{tabular}
    \caption{Comparison of the proposed VGG-regularized model and the baseline SR3 model. For the proposed method, we report the weights selected based on peak reconstruction quality and convergence, respectively.}
   \label{tab:vgg_best_l1_vs_l2_mri}
\end{table}
Although the peak PSNR and SSIM values are close, we observed that one of the selected VGG-regularized models ($\mathcal{L}^{(2)}_{\mathrm{VGG}}$, $\omega_{\mathrm{VGG}}=10$) has a clear advantage during the early stage of training. As shown in \Cref{fig:vgg_convergence_test}, PSNR and SSIM values of the proposed method increase more rapidly than those of the baseline SR3 model, with the largest difference occurring at the first evaluation checkpoint.

\begin{figure}[!htbp]
    \centering
    \begin{tabular}{cccc}
    LR  & HR & SR3 & Proposed \\
    \includegraphics[width=0.22\columnwidth]{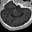}&
    \includegraphics[width=0.22\columnwidth]{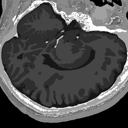}&
    \includegraphics[width=0.22\columnwidth]{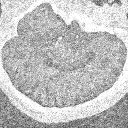}&
     \includegraphics[width=0.22\columnwidth]{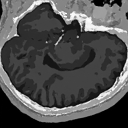}
  \end{tabular}
    \caption{BrainWeb MRI $4 \times$ reconstruction after 100K iterations. This corresponds to the first evaluation point (67 epochs) in \Cref{fig:vgg_convergence_test}. Our proposed method, shown in the rightmost panel, is obtained using $\mathcal{L}_{\mathrm{VGG}}^{(2)}$ with $\omega_{\mathrm{VGG}}=10$.}
    \label{fig:mri_early_reconstruction}
\end{figure}

To illustrate this difference, we compare the reconstructed images at
the early training stage. \Cref{fig:mri_early_reconstruction} shows the results
after 100K iterations, which correspond to the first evaluation point (67 epochs) in the convergence
curves in \Cref{fig:vgg_convergence_test}. At this stage, the SR3
model is still far from producing a stable reconstruction. In contrast, the
VGG-regularized model already recovers a much clearer brain structure, consistent with the faster convergence observed in \Cref{fig:vgg_convergence_test}.

\begin{figure}[!htbp]
    \centering
    \begin{tabular}{cccc}
    LR  & HR & SR3 & Proposed \\
    \includegraphics[width=0.22\columnwidth]{Brainweb_100k_100000_100_lr.png}&
    \includegraphics[width=0.22\columnwidth]{Brainweb_100k_100000_100_hr.png}&
    \includegraphics[width=0.22\columnwidth]{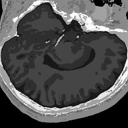}&
     \includegraphics[width=0.22\columnwidth]{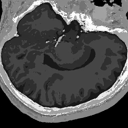}\\
        \includegraphics[width=0.22\columnwidth]{Brainweb_100k_100000_100_lr.png}&
    \includegraphics[width=0.22\columnwidth]{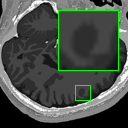}&
    \includegraphics[width=0.22\columnwidth]{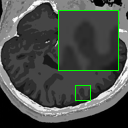}&
     \includegraphics[width=0.22\columnwidth]{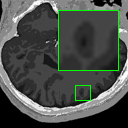}
  \end{tabular}
    \caption{BrainWeb MRI reconstruction after one million iterations, corresponding to the last evaluation point (667 epochs) in \Cref{fig:vgg_convergence_test}. The two rightmost panels show the reconstruction
from the proposed method ($\mathcal{L}_{\mathrm{VGG}}^{(2)}$,
$\omega_{\mathrm{VGG}}=10$). The zoomed regions show that the proposed method better preserves local structures and avoids connecting regions that should remain separated.}
    \label{fig:mri_late_reconstruction}
\end{figure}

We also compare the final reconstruction results after one million iterations, shown
in \Cref{fig:mri_late_reconstruction}. This corresponds to the last evaluation point (667 epochs) in
the convergence curves in \Cref{fig:vgg_convergence_test}. We observe both models can reconstruct the main brain structure, but the zoomed regions show that the VGG-regularized model preserves local shapes more accurately. In the baseline SR3 reconstruction, some nearby structures that should remain
separated become visually connected. By contrast, the proposed VGG-regularized SR3 reconstruction keeps these structures more clearly separated, suggesting that although the proposed method achieves similar final PSNR/SSIM as SR3, it improves perceptual quality and preserves fine details more effectively.

\subsubsection{BrainWeb MRI: results for the \texorpdfstring{$2\times$}{2x} and \texorpdfstring{$8\times$}{8x} tasks}

We next examine whether the early-convergence behavior observed in the $4\times$ task (\Cref{BrainWeb MRI: 4x}) persists at lower and higher super-resolution scale factors. The corresponding parameter sensitivities are discussed in \Cref{sec:para_sensitivity} as well.

\begin{figure}[!htbp]
\centering
\begin{subfigure}{0.47\linewidth}
\centering
\includegraphics[width=\linewidth]{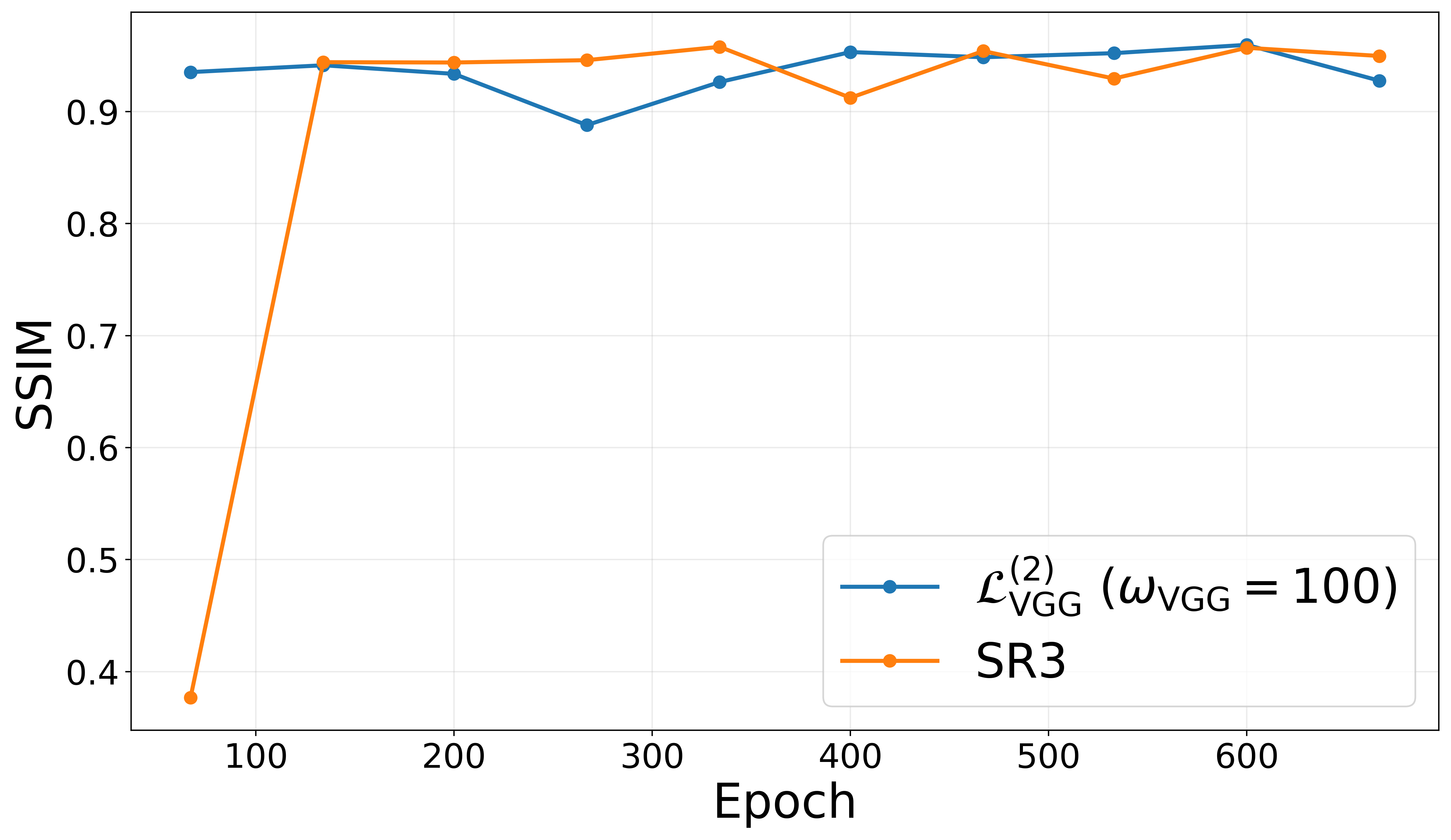}
\caption{SSIM}
\label{fig:brainweb_2x_ssim_convergence}
\end{subfigure}
\hfill
\begin{subfigure}{0.47\linewidth}
\centering
\includegraphics[width=\linewidth]{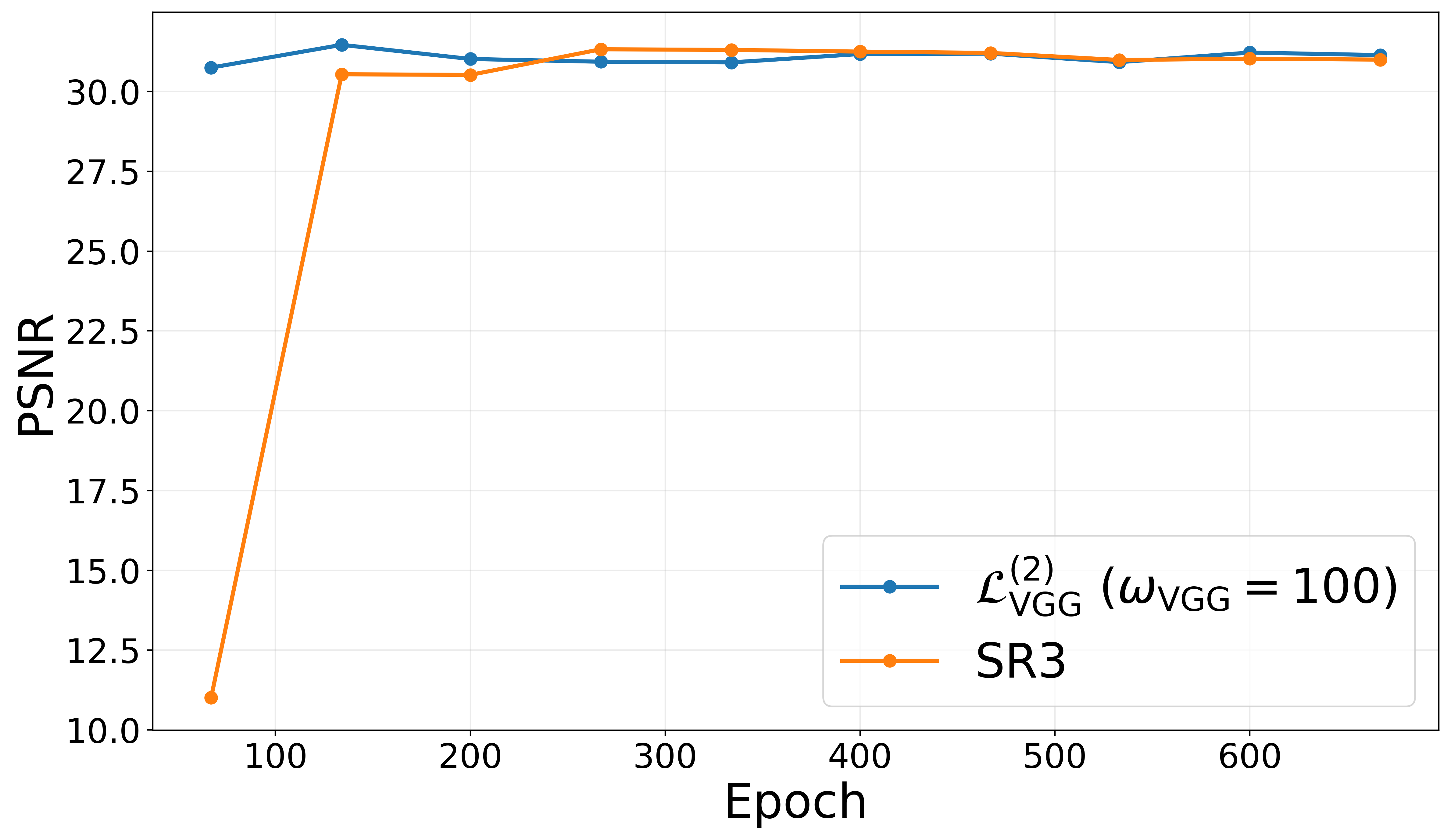}
\caption{PSNR}
\label{fig:brainweb_2x_psnr_convergence}
\end{subfigure}
\caption{Convergence comparison between the proposed method ($\mathcal{L}_{\mathrm{VGG}}^{(2)}$, $\omega_{\mathrm{VGG}}=100$; blue) and the baseline SR3 model (orange) on the BrainWeb $2\times$ task in terms of SSIM (left) and PSNR (right). The proposed method converges faster than SR3 and the two methods eventually reach similar SSIM and PSNR.}
\label{fig:brainweb_2x_convergence}
\end{figure}

\begin{figure}[!htbp]
    \centering
    \begin{tabular}{cccc}
    LR  & HR & SR3 & Proposed \\
    \includegraphics[width=0.22\columnwidth]{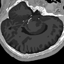}&
    \includegraphics[width=0.22\columnwidth]{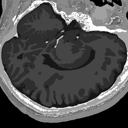}&
    \includegraphics[width=0.22\columnwidth]{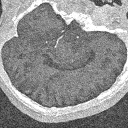}&
     \includegraphics[width=0.22\columnwidth]{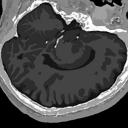}
  \end{tabular}
\caption{BrainWeb MRI $2 \times$ reconstruction after 100K iterations, corresponding to the
first evaluation point (67 epochs) in
\Cref{fig:brainweb_2x_convergence}. The rightmost panel shows the
reconstruction produced by the proposed method
$\mathcal{L}^{(2)}_{\mathrm{VGG}}$ with
$\omega_{\mathrm{VGG}}=100$.}
    \label{fig:mri_early_reconstruction_2x}
\end{figure}

For the $2\times$ task, the $\mathcal{L}^{(2)}_{\mathrm{VGG}}$ model with $\omega_{\mathrm{VGG}}=100$ is selected to compare with the baseline SR3 model. As shown in \Cref{fig:brainweb_2x_convergence}, the two models eventually reach similar PSNR and SSIM levels, but the VGG-regularized model performs much better at the first evaluation checkpoint. The difference is especially clear in the reconstruction at 100K iterations shown in \Cref{fig:mri_early_reconstruction_2x}: the SR3 output still contains noticeable noise, whereas the proposed VGG-regularized output is already close to the HR image.

\noindent\begin{minipage}{\linewidth}
\indent
For this $2\times$ setting, we omit the visual comparison at the final checkpoint
of one million iterations. Since the $2\times$ SR task is relatively easy, the
final reconstructions from SR3 and VGG-regularized SR3 are visually almost indistinguishable,
and no clear structural difference is observed. This is different from the
previous $4\times$ experiment and the next $8\times$ experiment, where the task
is more challenging and the final zoomed comparisons still reveal visible
differences in local details.
\end{minipage}

\begin{figure}[!htbp]
\centering
\begin{subfigure}{0.47\linewidth} \centering \includegraphics[width=\linewidth]{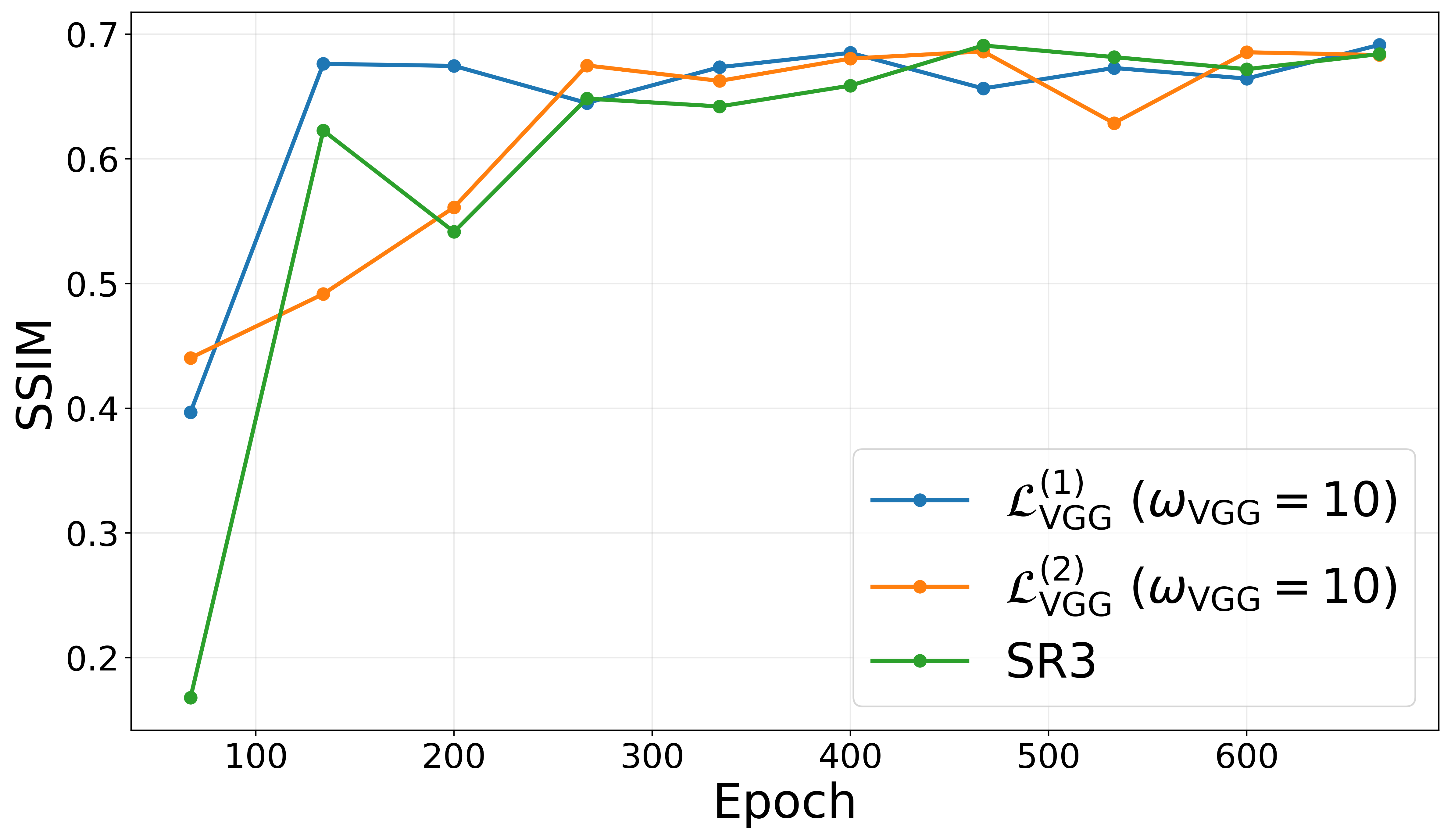} \caption{SSIM} \label{fig:brainweb_8x_psnr_convergence}
\end{subfigure}
\hfill
\begin{subfigure}{0.47\linewidth}
\centering \includegraphics[width=\linewidth]{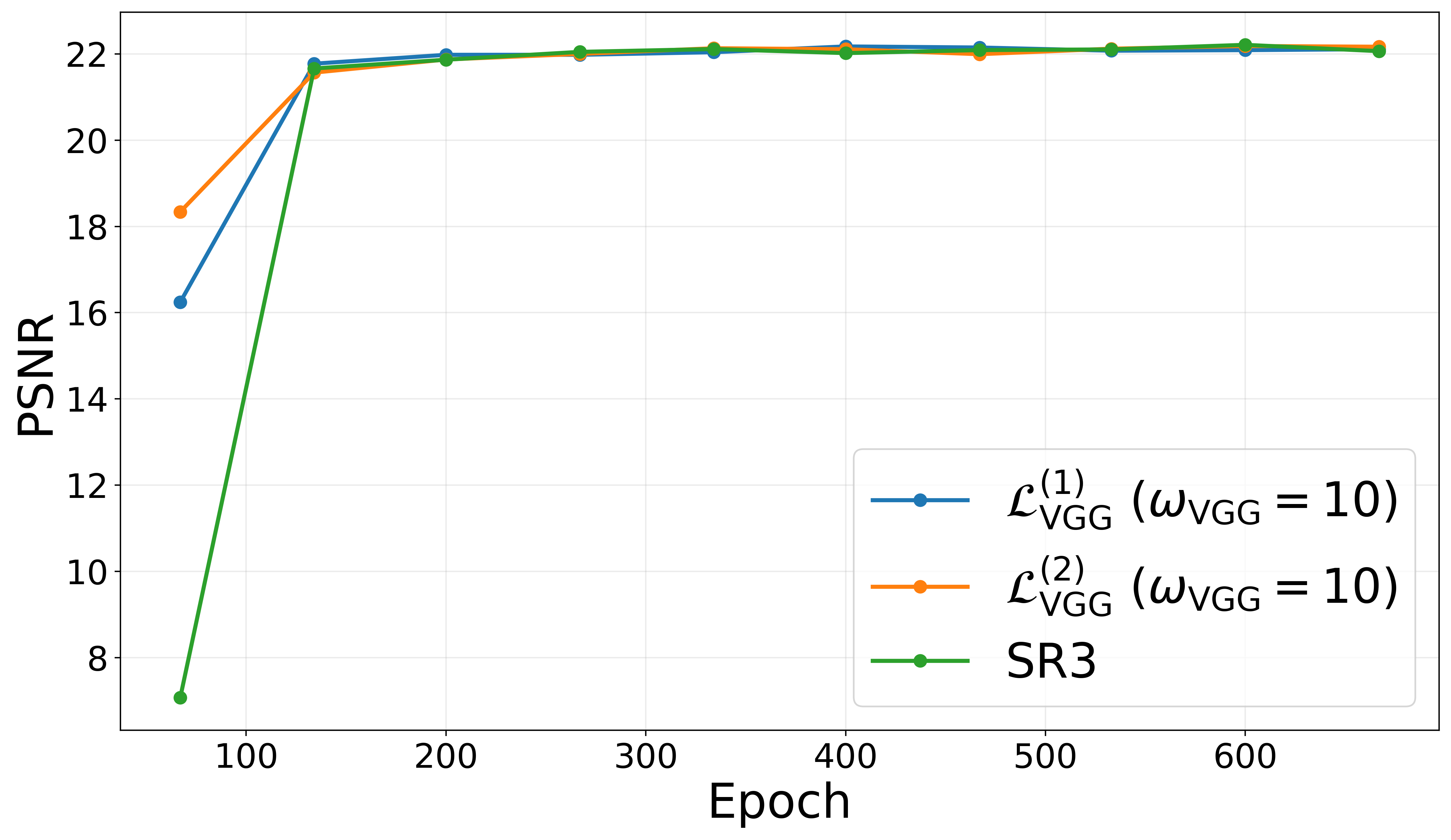}
\caption{PSNR}
 \end{subfigure}
\caption{Convergence comparison among the proposed VGG-regularized models using
$\mathcal{L}_{\mathrm{VGG}}^{(1)}$ with $\omega_{\mathrm{VGG}}=10$ (blue),
$\mathcal{L}_{\mathrm{VGG}}^{(2)}$ with $\omega_{\mathrm{VGG}}=10$ (orange),
and the baseline SR3 model (green) on the BrainWeb $8\times$ task. A faster convergence of the proposed models is observed.}
\label{fig:brainweb_8x_convergence}
\end{figure}

For the more challenging $8\times$ task, the convergence curves in \Cref{fig:brainweb_8x_convergence} again show that VGG-regularized models have faster early convergence. At the first evaluation checkpoint (100K iterations, 67 epochs), both selected VGG models already achieve higher PSNR and SSIM than SR3.

\begin{figure}[!htbp]
    \centering
    \begin{tabular}{cccc}
    LR  & HR & SR3 & Proposed \\
    \includegraphics[width=0.22\columnwidth]{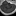}&
    \includegraphics[width=0.22\columnwidth]{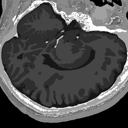}&
    \includegraphics[width=0.22\columnwidth]{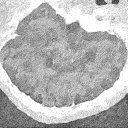}&
     \includegraphics[width=0.22\columnwidth]{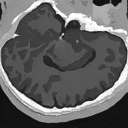}\\
  \end{tabular}
    \caption{BrainWeb MRI $8\times$ reconstruction after 100K iterations, corresponding to the first evaluation point (67 epochs) in \Cref{fig:brainweb_8x_convergence}. The rightmost panel shows the reconstruction produced by the proposed method using $\mathcal{L}_{\mathrm{VGG}}^{(2)}$ with $\omega_{\mathrm{VGG}}=10$.}
    \label{fig:brainweb_8x_convergence_image_compare}
\end{figure}

\begin{figure}[!htbp]
    \centering
    \begin{tabular}{cccc}
    LR  & HR & SR3 & Proposed \\
    \includegraphics[width=0.22\columnwidth]{Brainweb_100K_8x_100000_100_lr-2.png}&
    \includegraphics[width=0.22\columnwidth]{Brainweb_100K_8x_100000_100_hr-2.png}&
    \includegraphics[width=0.22\columnwidth]{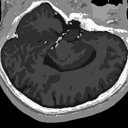}&
     \includegraphics[width=0.22\columnwidth]{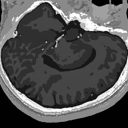}\\
        \includegraphics[width=0.22\columnwidth]{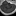}&
    \includegraphics[width=0.22\columnwidth]{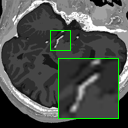}&
    \includegraphics[width=0.22\columnwidth]{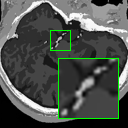}&
     \includegraphics[width=0.22\columnwidth]{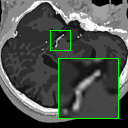}
  \end{tabular}
    \caption{BrainWeb MRI $8\times$ reconstruction after one million iterations. This corresponds to the last evaluation point (667 epochs) in \Cref{fig:brainweb_8x_convergence}. The rightmost panels show the reconstruction produced by the proposed method using $\mathcal{L}_{\mathrm{VGG}}^{(2)}$ with $\omega_{\mathrm{VGG}}=10$. The zoomed-in region highlights that our  method reconstructs the local structure more smoothly and continuously, more closely matching the HR reference, whereas the baseline SR3 reconstruction has a partially disconnected and irregular thickening appearance. }
    \label{fig:brainweb_8x_late_reconstruction}
\end{figure}

The image comparisons in Figures~\ref{fig:brainweb_8x_convergence_image_compare}
and \ref{fig:brainweb_8x_late_reconstruction} further illustrate the effect of VGG regularization on this challenging $8\times$ SR task. After 100K iterations (\Cref{fig:brainweb_8x_convergence_image_compare}), the
baseline SR3 output still contains a significant amount of noise. In contrast, the VGG-regularized ($\mathcal{L}^{(2)}_{\mathrm{VGG}}$, $\omega_{\mathrm{VGG}}=10$) reconstruction already shows a more recognizable brain boundary and much clearer details, which is consistent with its faster early convergence result. After one million iterations (\Cref{fig:brainweb_8x_late_reconstruction}), both
models recover the main structure, but the zoomed region shows that
VGG-regularized SR3 ($\mathcal{L}^{(2)}_{\mathrm{VGG}}$, $\omega_{\mathrm{VGG}}=10$) gives a cleaner reconstruction of the thin local structure. The baseline SR3 result is slightly more irregular in this region, while the VGG-regularized result better follows the shape of the HR image.

\subsubsection{Brain Tumor MRI: Results on the $4\times$ Task}

\begin{figure}[!htbp]
    \centering
    \begin{subfigure}{0.48\linewidth}
        \centering
        \includegraphics[width=\linewidth]{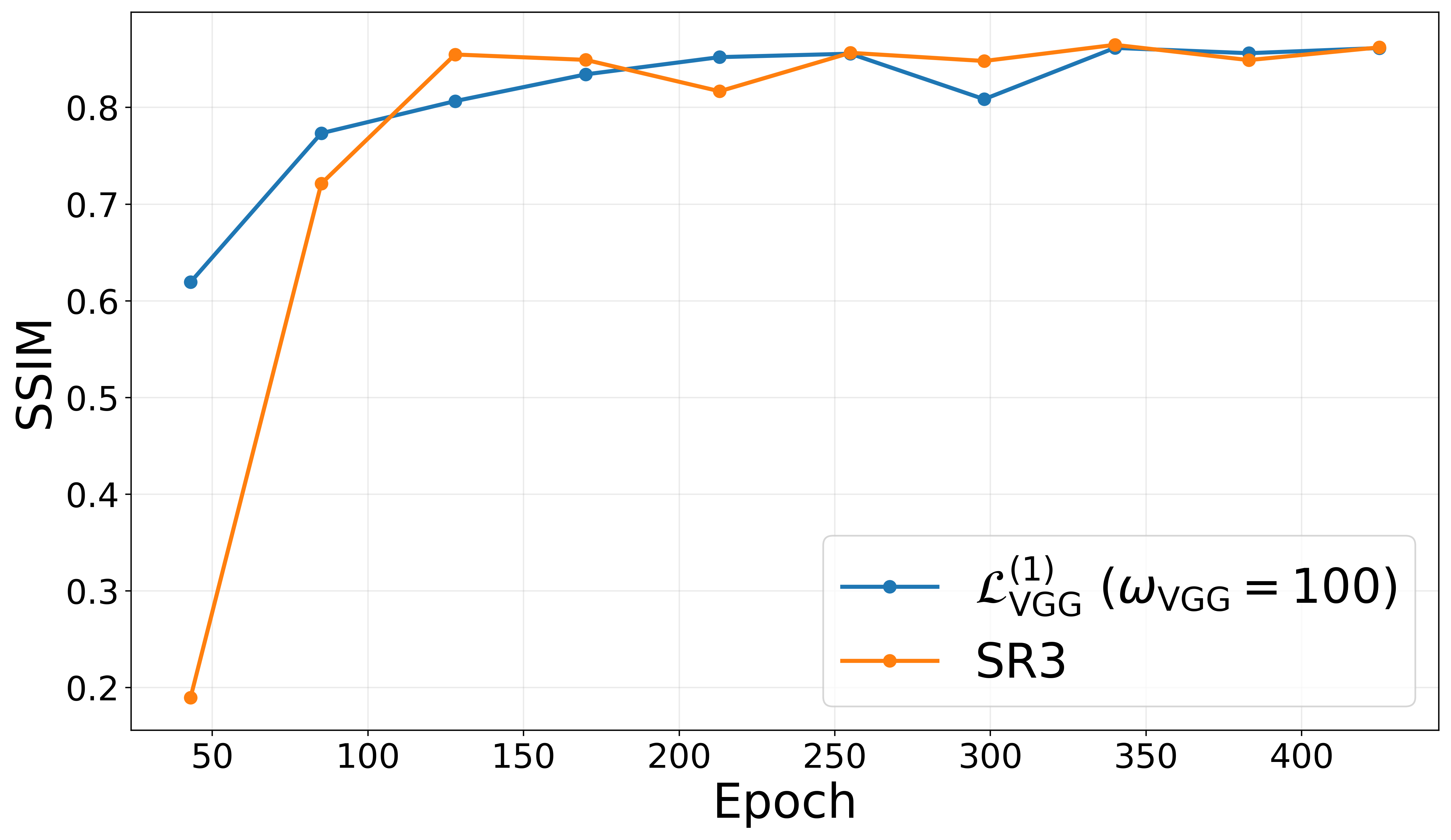}
        \caption{SSIM}
        \label{fig:tumor_4x_ssim_convergence}
    \end{subfigure}
    \hfill
    \begin{subfigure}{0.48\linewidth}
        \centering
        \includegraphics[width=\linewidth]{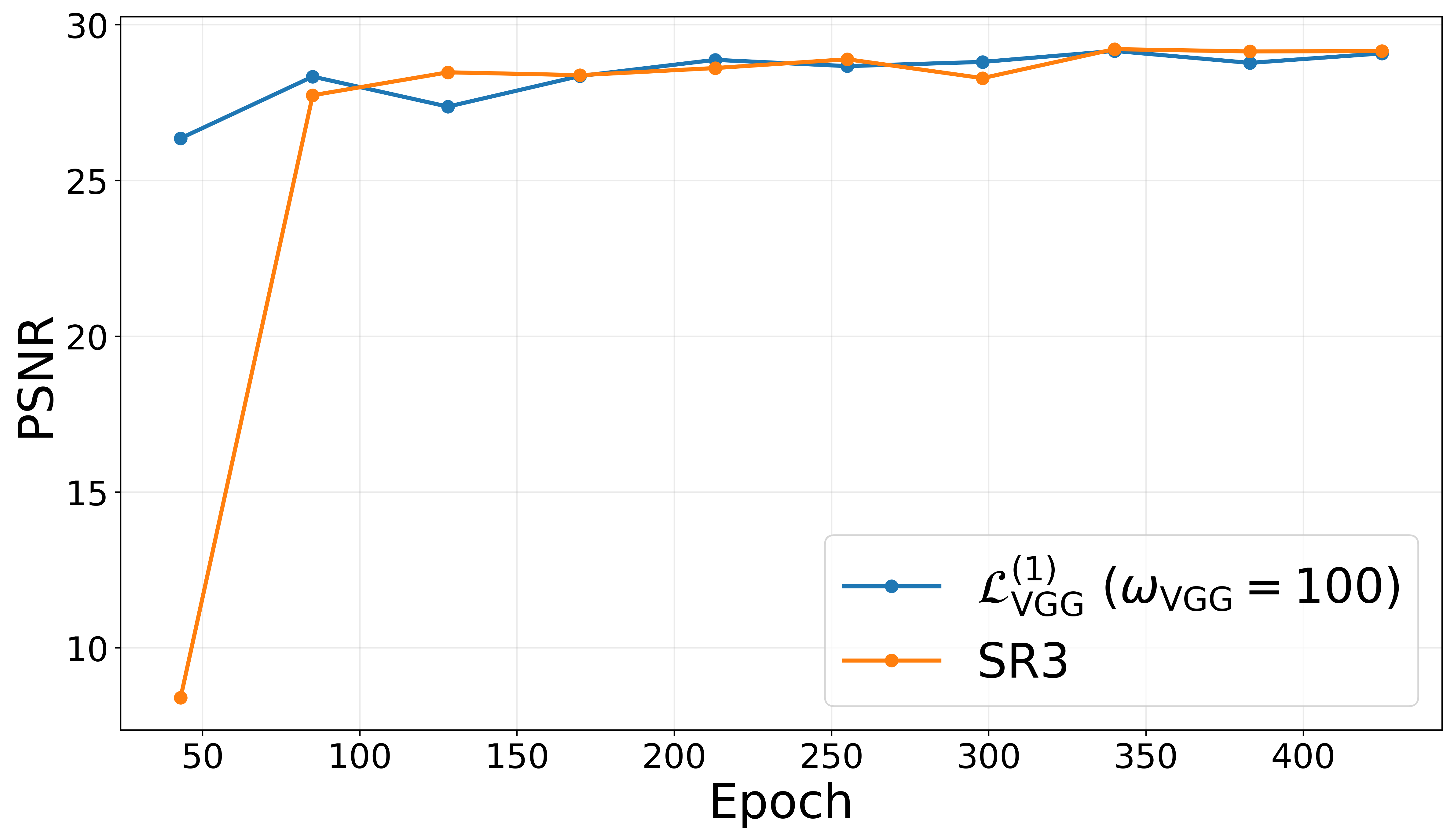}
        \caption{PSNR}
        \label{fig:tumor_4x_psnr_convergence}
    \end{subfigure}
\caption{Convergence comparison between the proposed method ($\mathcal{L}_{\mathrm{VGG}}^{(1)}$,
$\omega_{\mathrm{VGG}}=100$; blue) and the baseline SR3 model
(orange) on the Brain Tumor MRI $4\times$ task. The proposed method exhibits faster convergence.}
    \label{fig:tumor_4x_convergence}
\end{figure}

We further evaluate the proposed VGG-regularized SR3 models on the Brain Tumor MRI $4\times$ task.
The parameter-sensitivity results are discussed in \Cref{tab:tumor_4x_vgg_weight_comparison},  \Cref{sec:para_sensitivity}. For the convergence comparison, we use the $\mathcal{L}_{\mathrm{VGG}}^{(1)}$ model with $\omega_{\mathrm{VGG}}=100$, which provides the fastest early convergence among the tested parameters. As shown in \Cref{fig:tumor_4x_convergence}, the VGG-regularized model achieves much higher PSNR and SSIM at the first evaluation checkpoint. The difference gradually decreases as training proceeds, and the two models eventually reach similar PSNR/SSIM levels. This behavior is consistent with the BrainWeb experiments and suggests that the early-convergence benefit of VGG regularization is not specific to a single MRI dataset.

\begin{figure}[!htbp]
    \centering
    \begin{tabular}{cccc}
    LR  & HR & SR3 & Proposed \\
    \includegraphics[width=0.22\columnwidth]{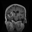}&
    \includegraphics[width=0.22\columnwidth]{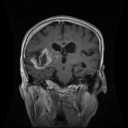}&
    \includegraphics[width=0.22\columnwidth]{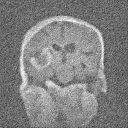}&
     \includegraphics[width=0.22\columnwidth]{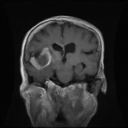}
  \end{tabular}
\caption{Brain Tumor MRI $4 \times$ reconstruction after 100K iterations, corresponding to the
first evaluation point (43 epochs) in
\Cref{fig:tumor_4x_convergence}. The rightmost panel shows the reconstruction
produced by the proposed method using
$\mathcal{L}_{\mathrm{VGG}}^{(1)}$ with
$\omega_{\mathrm{VGG}}=100$.}
    \label{Tumor_convergence_image_compare}
\end{figure}

\begin{figure}[!htbp]
    \centering
    \begin{tabular}{cccc}
    LR  & HR & SR3 & Proposed \\
    \includegraphics[width=0.22\columnwidth]{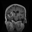}&
    \includegraphics[width=0.22\columnwidth]{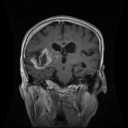}&
    \includegraphics[width=0.22\columnwidth]{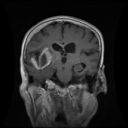}&
     \includegraphics[width=0.22\columnwidth]{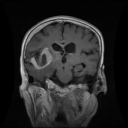}\\
        \includegraphics[width=0.22\columnwidth]{Tumor_1million_1000000_100_lr.png}&
    \includegraphics[width=0.22\columnwidth]{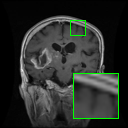}&
    \includegraphics[width=0.22\columnwidth]{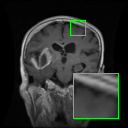}&
     \includegraphics[width=0.22\columnwidth]{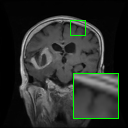}\\
        \includegraphics[width=0.22\columnwidth]{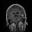}&
    \includegraphics[width=0.22\columnwidth]{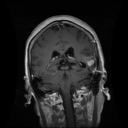}&
    \includegraphics[width=0.22\columnwidth]{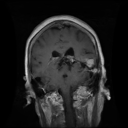}&
     \includegraphics[width=0.22\columnwidth]{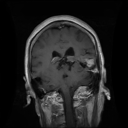}\\
        \includegraphics[width=0.22\columnwidth]{Tumor_1million_1000000_17_lr.png}&
    \includegraphics[width=0.22\columnwidth]{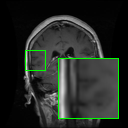}&
    \includegraphics[width=0.22\columnwidth]{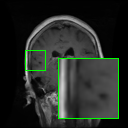}&
     \includegraphics[width=0.22\columnwidth]{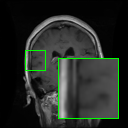}
  \end{tabular}
    \caption{Brain Tumor MRI reconstruction after one million iterations, corresponding to
the last evaluation point (425 epochs) in
\Cref{fig:tumor_4x_convergence}. The rightmost panels show the reconstructions
produced by the proposed method using
$\mathcal{L}_{\mathrm{VGG}}^{(1)}$ with
$\omega_{\mathrm{VGG}}=100$. The zoomed-in regions show that the proposed
method better preserves local boundaries and structural contrast. }
    \label{fig:Tumor_mri_late_reconstruction}
\end{figure}

The visual comparison for Brain Tumor MRI SR task shows a similar early-stage advantage of VGG regularization to that observed for the Brainweb MRI. After 100K iterations, the SR3 reconstruction is still dominated by noise. The proposed VGG-regularized SR3, however, already reconstructs a much more stable image, with clearer brain contours and a more clearly defined local boundary, see \Cref{Tumor_convergence_image_compare}. After one million iterations (\Cref{fig:Tumor_mri_late_reconstruction}), the global reconstructions from SR3 and proposed VGG-regularized SR3 become very close, but the zoomed regions show remaining local differences. In particular, the proposed VGG-regularized SR3 better preserves the edge contrast, while the baseline SR3 result appears slightly more blurred and misses some local structural details.

Together with the BrainWeb results, these comparisons indicate that for grayscale MRI images, VGG regularization consistently improves early-stage reconstruction and can also preserve local structural details in the final reconstruction.

\subsection{Color Face Image Super-Resolution}
\label{subsec:color_celeb}

\begin{figure}[!htbp]
    \centering
    \begin{subfigure}{0.49\linewidth}
        \centering
        \includegraphics[width=\linewidth]{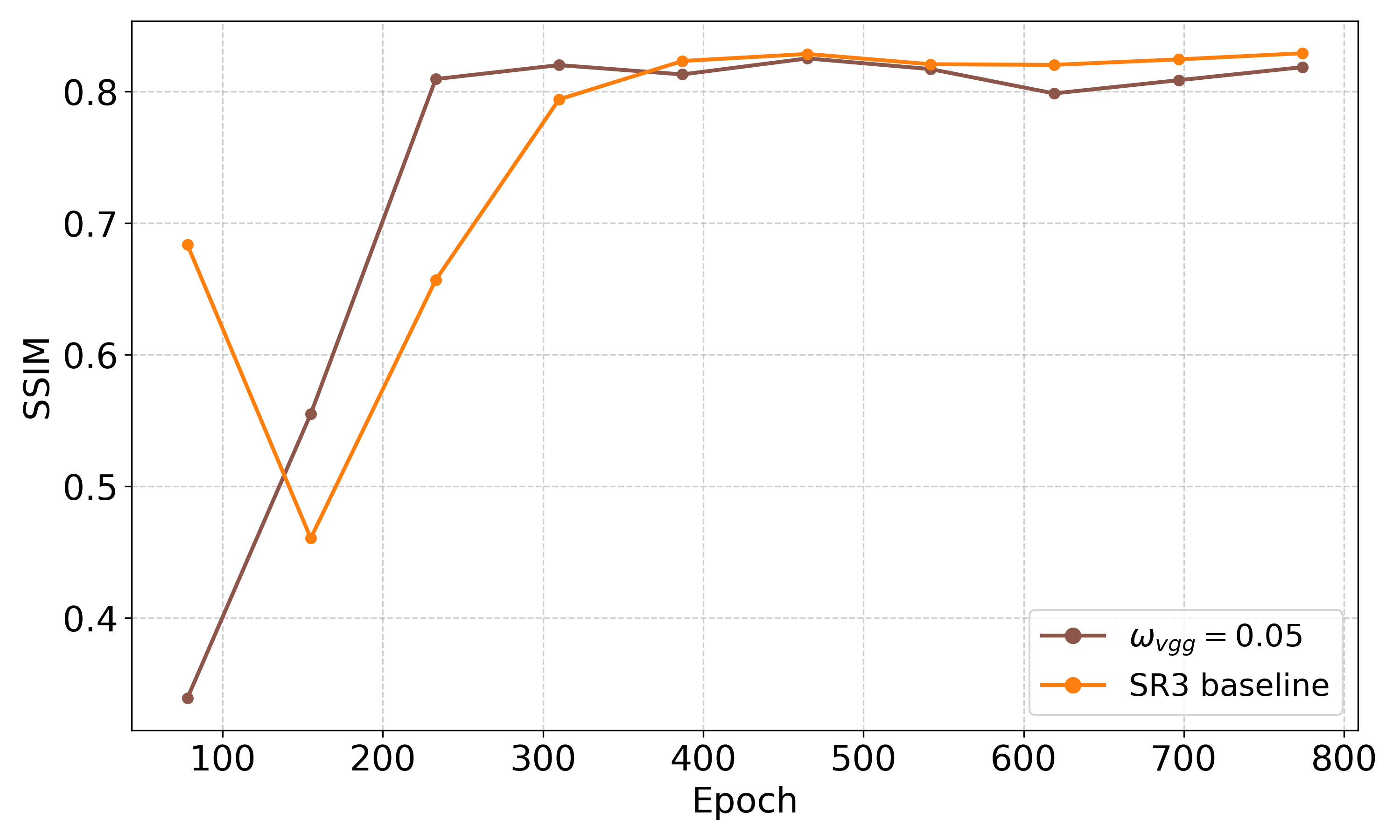}
        \caption{SSIM}
    \end{subfigure}
    \hfill
    \begin{subfigure}{0.49\linewidth}
        \centering
        \includegraphics[width=\linewidth]{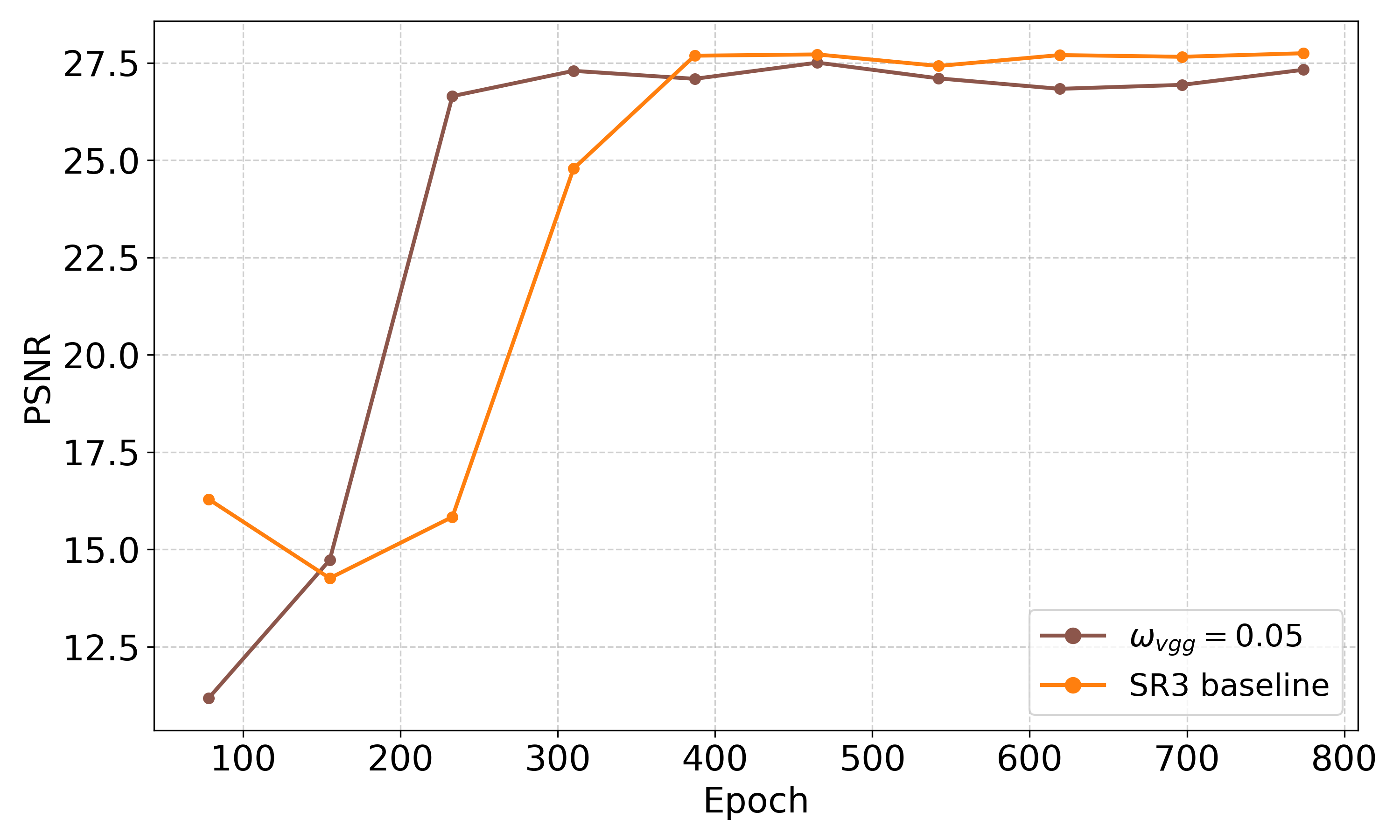}
        \caption{PSNR}
    \end{subfigure}
    \caption{Comparison of PSNR and SSIM values between the SR3 baseline (orange) and the proposed method using
$\mathcal{L}_{\mathrm{VGG}}^{(1)}$ with
$\omega_{\mathrm{VGG}}=0.05$ (brown). Our method reaches a stable performance level by approximately epoch 233, substantially earlier than the SR3 baseline. This indicates that VGG loss stabilizes early training and accelerates optimization, though this weight's peak PSNR/SSIM trails the baseline slightly, even as it yields cleaner visual reconstructions (Figure~\ref{fig:celeb_zoomed_stack_005}).}\label{fig:celeb_vgg}
\end{figure}

\begin{figure}[!htbp]
    \centering
    \setlength{\tabcolsep}{1pt}
    \begin{tabular}{cccc}
        \small LR & \small HR & \small SR3 & \small Proposed \\
        \includegraphics[width=0.24\columnwidth]{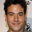} &
        \includegraphics[width=0.24\columnwidth]{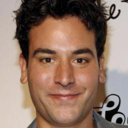} &
        \includegraphics[width=0.24\columnwidth]{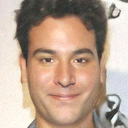} &
        \includegraphics[width=0.24\columnwidth]{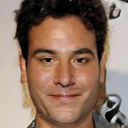} \\
    \end{tabular}
 \caption{Visual comparison of $4\times$ facial reconstruction on CelebA-HQ at epoch 233 in Fig.~\ref{fig:celeb_vgg}. From left to right: Low-Resolution (LR) input, High-Resolution (HR) ground truth, the SR3 baseline, and our proposed method. The SR3 baseline exhibits a washed-out, overexposed artifact that distorts natural skin tones. In contrast, the proposed VGG variant maintains proper color depth, contrast, and structural definition early in the training cycle.}\label{fig:celeb_firstepoch}
\end{figure}

\begin{figure}[!htbp]
    \centering
    \begin{tabular}{c}
            \begin{tabular}{>{\centering\arraybackslash}p{0.22\columnwidth} >{\centering\arraybackslash}p{0.22\columnwidth} >{\centering\arraybackslash}p{0.22\columnwidth} >{\centering\arraybackslash}p{0.22\columnwidth}}
            \small LR & \small HR & \small SR3 & \small Proposed \\
        \end{tabular} \\
        \includegraphics[width=0.98\columnwidth]{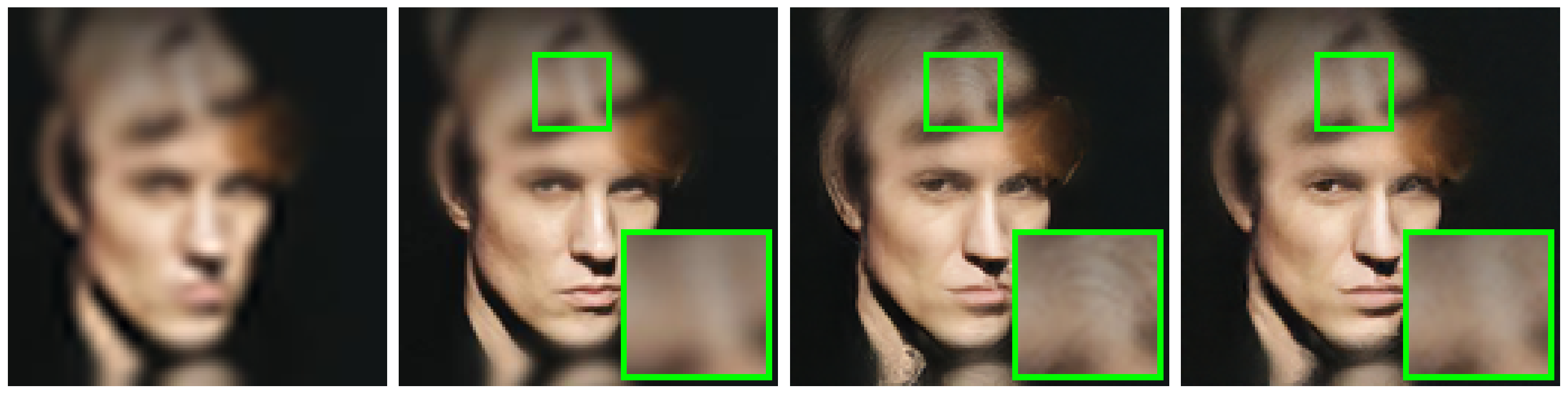} \\
        \includegraphics[width=0.98\columnwidth]{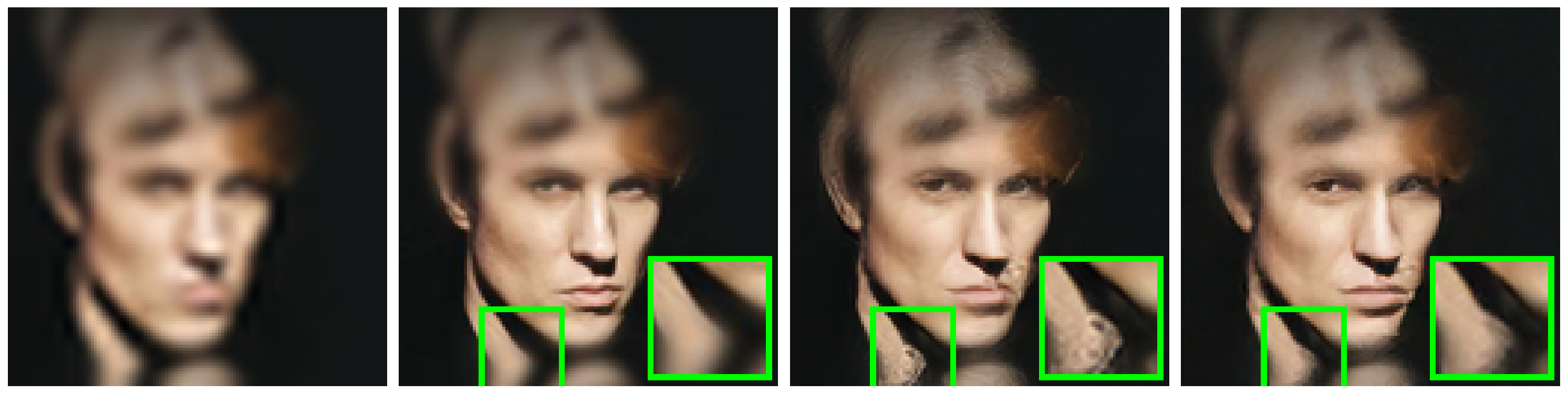} \\
    \end{tabular}
\caption{Visual comparison of $4\times$ facial reconstruction on CelebA-HQ at the last evaluation point in Fig.~\ref{fig:celeb_vgg}. From left to right: Low-Resolution (LR) input, High-Resolution (HR) ground truth, SR3 baseline, and our proposed model ($\omega_{\mathrm{VGG}} = 0.05$). Green boxes show zoomed details of the neck/collar boundary (top row) and the hairline and cheek region (bottom row). The proposed model produces smoother, more continuous boundaries at the collar than the SR3 baseline, and shows a more modest reduction in fine-texture noise around the hairline.}
\label{fig:celeb_zoomed_stack_005}
\end{figure}

{We next test the generalizability of the proposed method on color image super-resolution with training and testing done on different datasets. All color RGB configurations evaluate the  super-resolution task at a single scale factor ($4\times$) using the proposed VGG-regularized SR3 model  $\mathcal{L}^{(1)}_{VGG}$. The training is done on the FFHQ dataset and the testing is done on the CelebA-HQ dataset. The sensitivity analysis used to select the optimal VGG loss weight is detailed separately in Section~\ref{sec:para_sensitivity}.} As in the grayscale MRI experiments in Section~\ref{subsec:gray_mri}, the weight offering the fastest early convergence does not necessarily yield the best final reconstruction quality. Figures~\ref{fig:celeb_vgg}– ~\ref{fig:celeb_zoomed_stack_005} use $\omega_{\mathrm{VGG}} = 0.05$ to illustrate the convergence advantage of perceptual regularization; the finer-grained sweep in Section~\ref{sec:para_sensitivity}  identifies $\omega_{\mathrm{VGG}} = 0.04$ as the setting with the best overall final PSNR and SSIM.

Figure~\ref{fig:celeb_vgg} illustrates the evaluation metrics across 800 training epochs. Incorporating the perceptual loss (\(\omega_{\mathrm{VGG}} = 0.05\)) significantly accelerates convergence relative to the standard SR3 baseline. At epoch 230, the VGG-regularized model achieves a PSNR of approximately 26.6 dB and an SSIM of 0.81, closely approaching its asymptotic performance. In contrast, the baseline requires roughly 390 epochs to reach an equivalent performance threshold, representing an approximate 40\% reduction in necessary training iterations to achieve stability. Beyond convergence velocity, the perceptual weight stabilizes early-stage optimization. Between epochs 75 and 150, the baseline exhibits a distinct performance degradation in both metrics, indicating unstable early-stage optimization before identifying a viable optimization path. The VGG variant mitigates this issue, maintaining a monotonic upward trajectory from the onset of training. This particular weight trades some final-epoch quality for faster convergence.

 A qualitative analysis in Figure~\ref{fig:celeb_firstepoch} reveals distinct advantages in the color and contrast preservation of the proposed method over the baseline at epoch 233. While the standard SR3 baseline yields a severely washed-out reconstruction with overexposed, unnaturally pale skin tones, the VGG-regularized model maintains proper color depth and contrast. By anchoring optimization within a perceptual feature space, the proposed method effectively mitigates early-stage luminance degradation and chalky artifacts, yielding a much closer match to the natural color profile of the target domain (HR).

Figure~\ref{fig:celeb_zoomed_stack_005} visually compares the models at the final evaluation checkpoint. In the first row, the zoomed region highlights the hairline. The standard SR3 baseline generates severe high-frequency noise and unnatural checkerboard textures across the skin and hair strands. In contrast, our proposed model ($\omega_{\mathrm{VGG}} = 0.05$) completely suppresses these noisy artifacts, producing smooth skin surfaces and realistic hair shading that closely match the high-resolution ground truth. In the second row, the zoomed region focuses on the sharp edge of the  neck. The SR3 baseline fails to form a clean boundary, creating jagged, blurred edges and visible pixel noise. Our proposed framework successfully removes these distortions, restoring a sharp, smooth edge transition. These visual results confirm that regularizing the model with a VGG loss effectively corrects localized structural artifacts that standard pixel-level training fails to fix.

\subsection{Discussions}

\subsubsection{Sensitivity and Parameter Selection}\label{sec:para_sensitivity}
We discuss the sensitivity of the numerical results to the VGG regularization parameters. Because the grayscale MRI and RGB facial-image experiments use different datasets, loss formulations, and values of $\omega_{\mathrm{VGG}}$, their numerical results are not compared directly. Instead, parameter sensitivity is analyzed separately within each experimental setting, followed by a discussion of the trends shared across the two image domains.

\paragraph{Grayscale MRI experiments.}
The effect of VGG regularization depends on both the perceptual loss type and the weight assigned to it. The following sensitivity discussion of the grayscale MRI experiments highlights three main observations. First, the VGG regularization parameters that yield the fastest early-stage convergence in \Cref{subsec:gray_mri}  usually do not produce the highest peak PSNR or SSIM. Second, the preferred parameters are not directly transferable across different super-resolution scale factors, even when the same dataset is used. Third, the preferred parameters are also not directly transferable across datasets, even when both datasets contain brain MRI images, as in our BrainWeb MRI and Brain Tumor MRI experiments.

For each grayscale MRI task, we evaluate the $\mathcal{L}_{\mathrm{VGG}}^{(1)}$ loss \eqref{eq:VGG_Loss_L1} and the $\mathcal{L}^{(2)}_{\mathrm{VGG}}$ loss \eqref{eq:VGG_Loss_L2} using
\begin{equation} \omega_{\mathrm{VGG}} \in \{0.1,1,10,100,1000\}.
\end{equation}
This sensitivity analysis focuses on parameter selection according to the peak PSNR and SSIM attained. As shown in \Cref{subsec:gray_mri}, the setting with the highest peak PSNR/SSIM does not necessarily yield the fastest early convergence. We demonstrate this distinction using the BrainWeb $4\times$ task below. For the remaining grayscale MRI tasks, we will focus only on the peak PSNR and SSIM used for parameter selection, as their early-convergence behavior and corresponding configurations selected have already been presented in \Cref{subsec:gray_mri}.

\begin{figure}[!htbp]
    \centering
    \begin{subfigure}{0.49\linewidth}
        \centering
        \includegraphics[width=\linewidth]{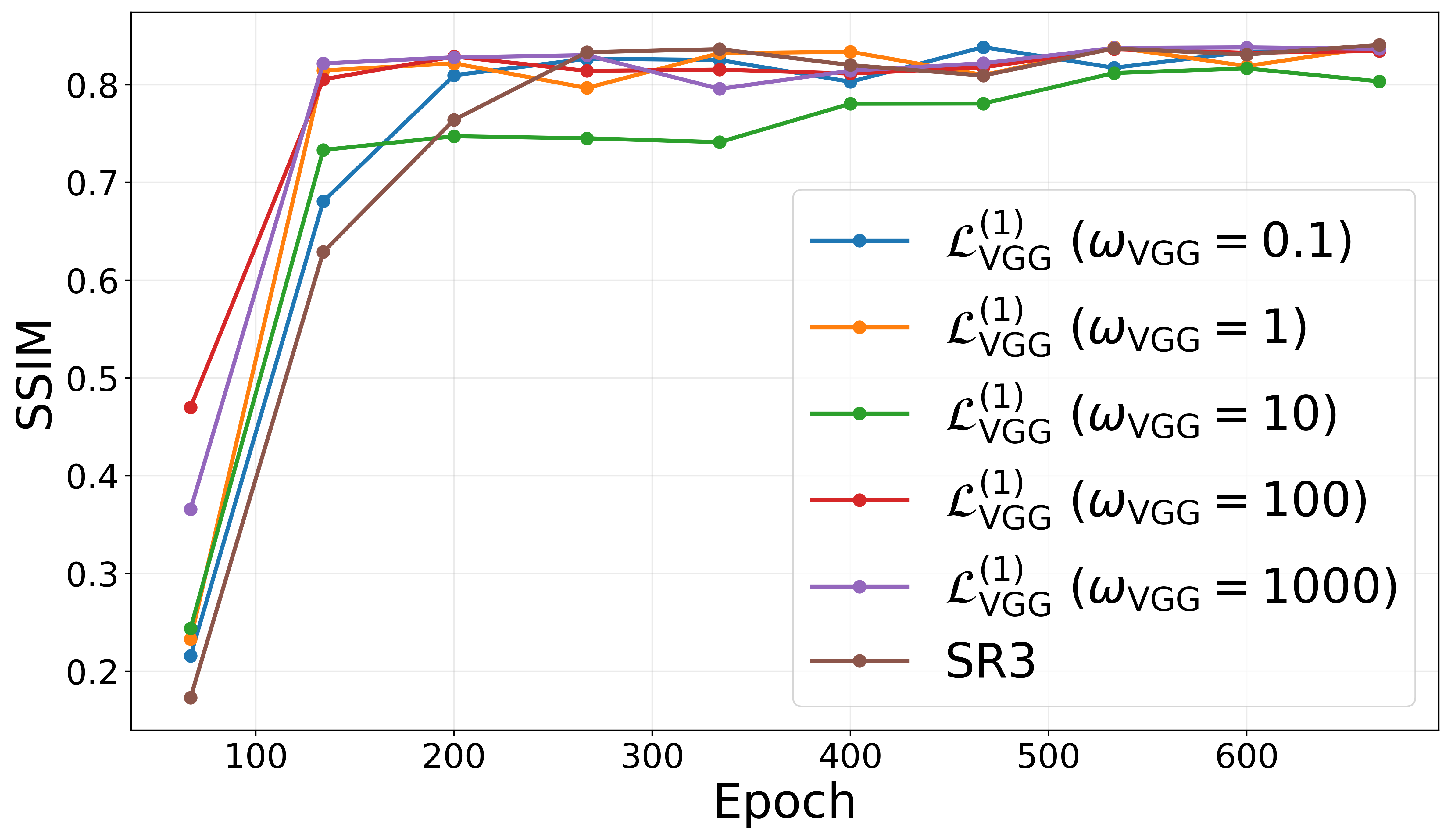}
        \caption{SSIM}
        \label{fig:vgg_ssim_mri}
    \end{subfigure}
    \hfill
    \begin{subfigure}{0.49\linewidth}
        \centering
        \includegraphics[width=\linewidth]{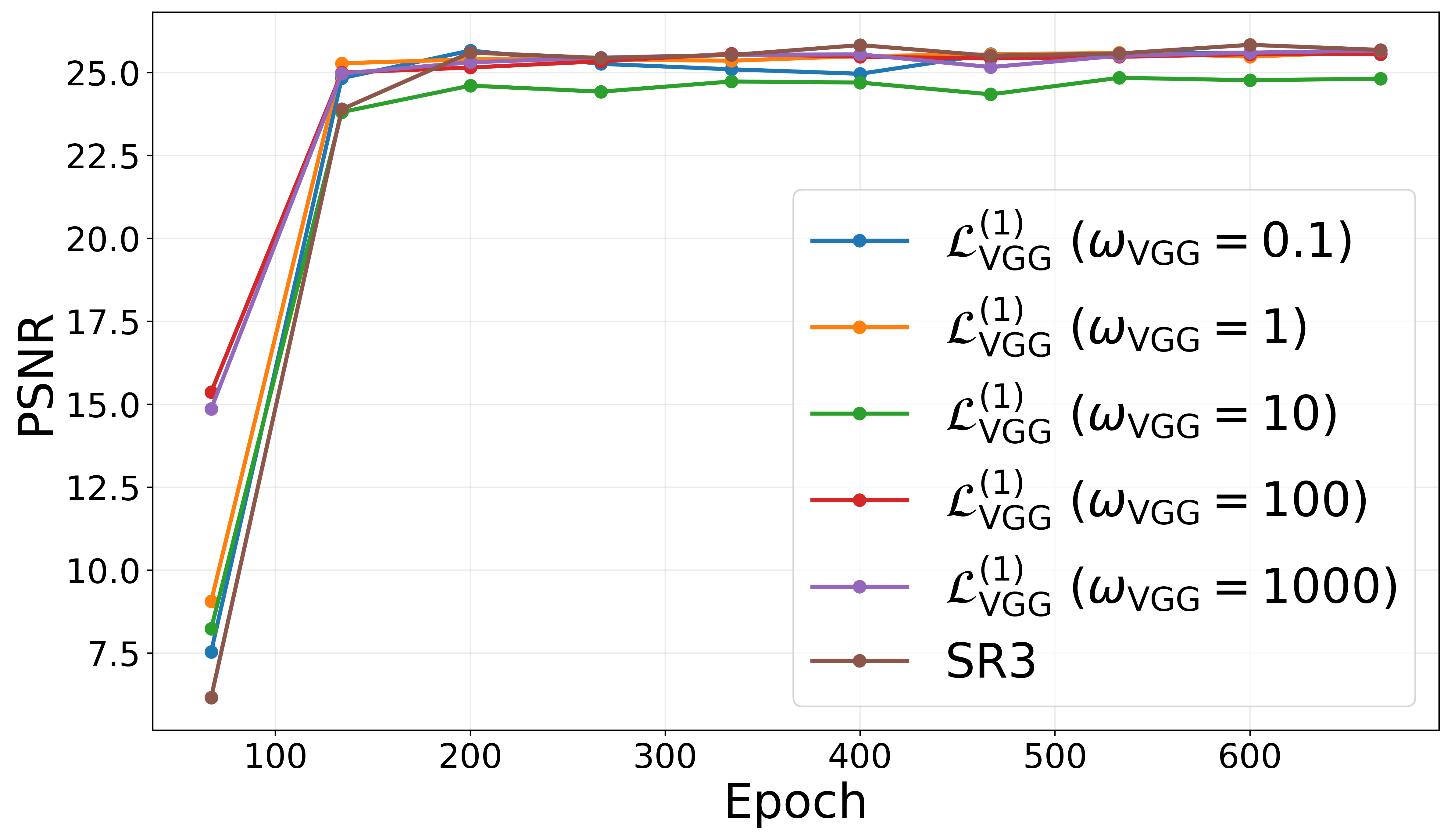}
        \caption{PSNR}
        \label{fig:vgg_psnr_mri}
    \end{subfigure}
    \caption{Parameter tuning results for  $\mathcal{L}_{\mathrm{VGG}}^{(1)}$ loss on BrainWeb MRI $4 \times$ SR. The model with $\omega_{\mathrm{VGG}}=100$ shows the fastest early convergence.}    \label{fig:vgg_l1_parameter_tune_mri}
\end{figure}

We begin with the BrainWeb $4\times$ task, which provides the most detailed sensitivity analysis. For the $\mathcal{L}_{\mathrm{VGG}}^{(1)}$ loss, the convergence curves in \Cref{fig:vgg_l1_parameter_tune_mri} show $\omega_{\mathrm{VGG}}={100}$ reaches a stable PSNR and SSIM level fastest in the early epochs.

However, \Cref{tab:vgg_weight_comparison_mri} shows that the best overall performance is obtained when $\omega_{\mathrm{VGG}}=0.1$. Therefore, for the $\mathcal{L}_{\mathrm{VGG}}^{(1)}$ loss, $\omega_{\mathrm{VGG}}=100$ is the best choice from the perspective of early convergence, while $\omega_{\mathrm{VGG}}=0.1$ gives the slightly better final reconstruction quality.

\begin{table}[!htbp]
\centering
\begin{tabular}{c|cc|cc}
\hline
\multirow{2}{*}{$\omega_{\mathrm{VGG}}$}
& \multicolumn{2}{c|}{$\mathcal{L}_{\mathrm{VGG}}^{(1)}$}
& \multicolumn{2}{c}{$\mathcal{L}_{\mathrm{VGG}}^{(2)}$} \\
& Peak PSNR & Peak SSIM & Peak PSNR & Peak SSIM \\
\hline
$0.1$  & \textbf{25.665} & 0.8384 & 25.452 & 0.8365 \\
$1$    & 25.632 & \textbf{0.8390} & \textbf{25.753} & \textbf{0.8436} \\
$10$   & 24.844 & 0.8166 & 25.660 & 0.8365 \\
$100$  & 25.569 & 0.8364 & 25.601 & 0.8366 \\
$1000$ & \textbf{25.665} & 0.8380 & 24.900 & 0.8169 \\
\hline
\end{tabular}
\caption{Parameter tuning of the VGG perceptual-loss weight $\omega_{\mathrm{VGG}}$ on BrainWeb MRI $4\times$ super-resolution. Results are reported for $\mathcal{L}_{\mathrm{VGG}}^{(1)}$ (left) and $\mathcal{L}_{\mathrm{VGG}}^{(2)}$ (right). The best values for each metric are highlighted in boldface.}
\label{tab:vgg_weight_comparison_mri}
\end{table}

\begin{figure}[!htbp]
    \centering
    \begin{subfigure}{0.48\linewidth}
        \centering
        \includegraphics[width=\linewidth]{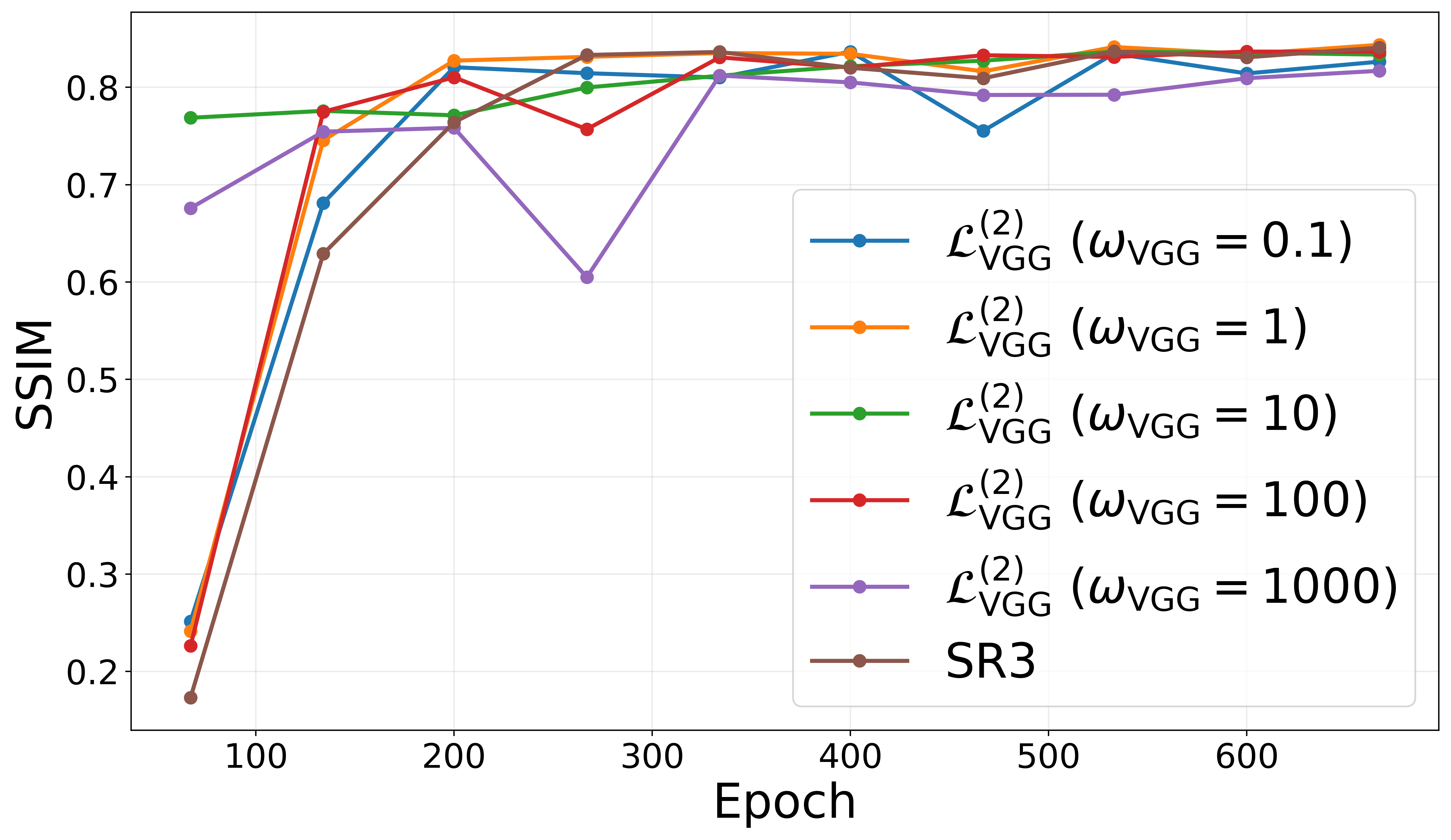}
        \caption{SSIM}
        \label{fig:vgg_ssim_mri_l2}
    \end{subfigure}
    \hfill
    \begin{subfigure}{0.48\linewidth}
        \centering
        \includegraphics[width=\linewidth]{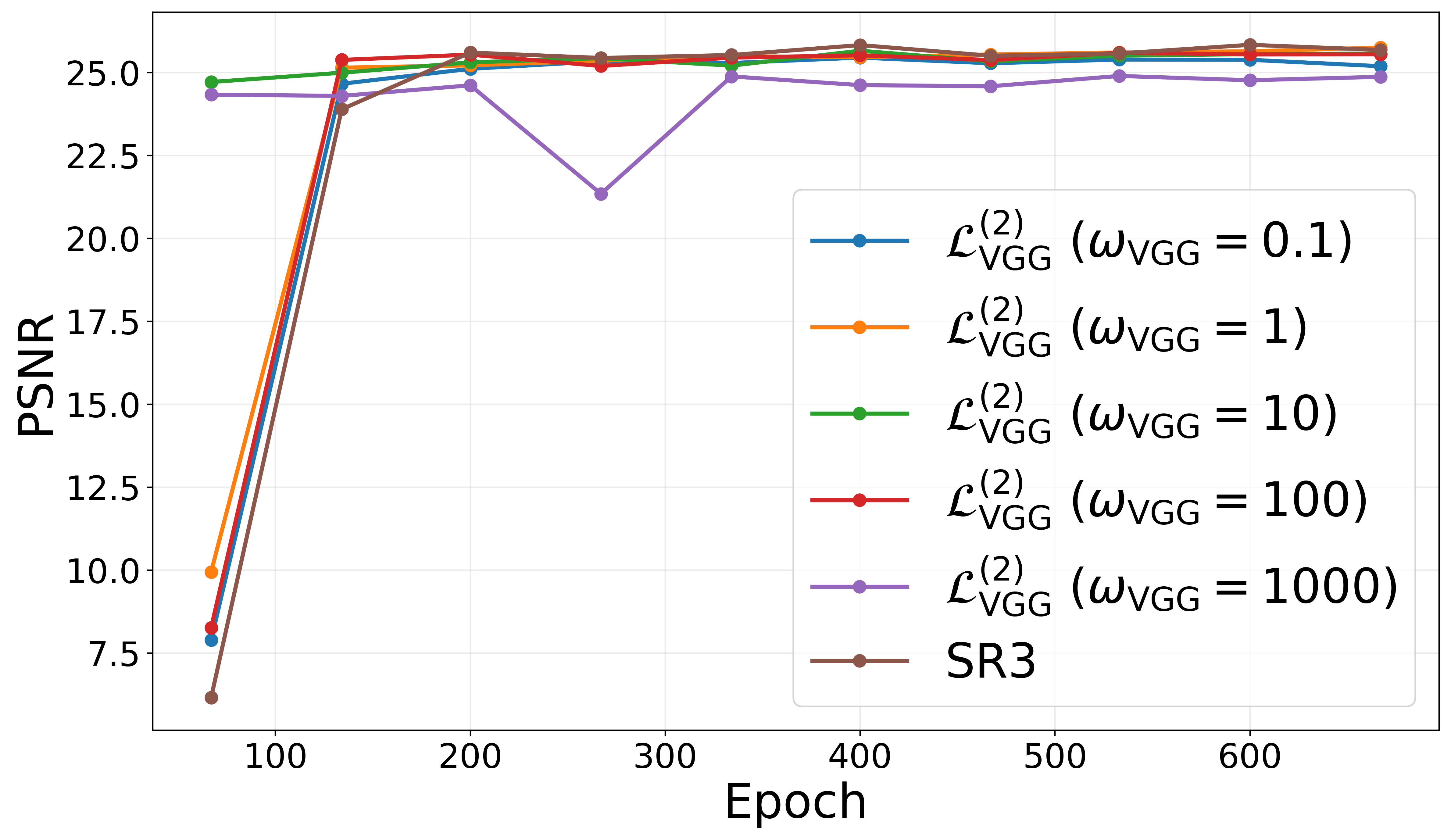}
        \caption{PSNR}
        \label{fig:vgg_psnr_mri_l2}
    \end{subfigure}
    \caption{Parameter tuning results for $\mathcal{L}^{(2)}_{\mathrm{VGG}}$ loss on BrainWeb MRI $4 \times$ SR. $\omega_\mathrm{VGG} = 10$ shows the fastest early convergence in both PSNR and SSIM.}
    \label{fig:vgg_l2_parameter_tune_mri}
\end{figure}

The $\mathcal{L}^{(2)}_{\mathrm{VGG}}$ loss exhibits a similar separation between the two criteria. As shown in \Cref{fig:vgg_l2_parameter_tune_mri} and \Cref{tab:vgg_weight_comparison_mri}, $\omega_{\mathrm{VGG}}=10$ gives the fastest early convergence in terms of both PSNR and SSIM. On the other hand, the best final reconstruction performance is achieved when $\omega_{\mathrm{VGG}}=1$. We also observe that an overly large VGG weight, such as $\omega_{\mathrm{VGG}}=1000$, leads to a clear drop in both PSNR and SSIM, suggesting that too much emphasis on the perceptual loss term can hurt the reconstruction quality.

The results for the remaining BrainWeb $2\times$ and $8\times$ tasks show that the preferred VGG configuration depends on the super-resolution scale factor. Although all three tasks use the same BrainWeb dataset, the parameters selected for the $2\times$ and $8\times$ tasks differ from those selected for the $4\times$ task. For the easier BrainWeb $2\times$ SR task, we repeat the same sensitivity study as the above $4\times$ SR task. The results are summarized in \Cref{tab:brainweb_2x_vgg_weight_comparison}. Since the $2\times$ task is less challenging than the $4\times$ and $8\times$ cases, all models achieve high PSNR and SSIM values. In particular, the $\mathcal{L}_{\mathrm{VGG}}^{(1)}$ model with $\omega_{\mathrm{VGG}}=100$ gives the best PSNR, while the $\mathcal{L}^{(2)}_{\mathrm{VGG}}$ model with $\omega_{\mathrm{VGG}}=100$ gives the best SSIM.

\begin{table}[!htbp]
\centering
\begin{tabular}{c|cc|cc}
\hline
\multirow{2}{*}{$\omega_{\mathrm{VGG}}$}
& \multicolumn{2}{c|}{$\mathcal{L}_{\mathrm{VGG}}^{(1)}$}
& \multicolumn{2}{c}{$\mathcal{L}_{\mathrm{VGG}}^{(2)}$} \\
& Peak PSNR & Peak SSIM & Peak PSNR & Peak SSIM \\
\hline
0.1   & 31.367 & 0.9584 & 31.216 & 0.9574 \\
1     & 31.226 & 0.9591 & 31.434 & 0.9595 \\
10    & 31.284 & 0.9581 & 31.269 & 0.9591 \\
100   & \textbf{31.605} & {0.9596} & {31.463} & \textbf{0.9597} \\
1000  & 31.245 & 0.9579 & 31.392 & 0.9583 \\
\hline
SR3 & 31.324 & 0.9579
    & 31.324 & 0.9579 \\
\hline
\end{tabular}
\caption{Sensitivity study of the VGG loss weight on the BrainWeb $2\times$ task. The $\mathcal{L}_{\mathrm{VGG}}^{(1)}$ model with $\omega_{\mathrm{VGG}}=100$ achieves the best PSNR, while the $\mathcal{L}^{(2)}_{\mathrm{VGG}}$ model with $\omega_{\mathrm{VGG}}=100$ achieves the best SSIM.}\label{tab:brainweb_2x_vgg_weight_comparison}
\end{table}

For the more challenging BrainWeb $8\times$ task, \Cref{tab:brainweb_8x_vgg_weight_comparison} shows VGG-regularized models achieve better numerical performance in terms of both PSNR and SSIM. In particular, the $\mathcal{L}_{\mathrm{VGG}}^{(1)}$ model with $\omega_{\mathrm{VGG}}=100$ gives the highest PSNR, while the $\mathcal{L}^{(2)}_{\mathrm{VGG}}$ model with $\omega_{\mathrm{VGG}}=100$ gives the highest SSIM.

\begin{table}[htbp]
\centering
\begin{tabular}{c|cc|cc}
\hline
\multirow{2}{*}{$\omega_{\mathrm{VGG}}$}
& \multicolumn{2}{c|}{$\mathcal{L}_{\mathrm{VGG}}^{(1)}$}
& \multicolumn{2}{c}{$\mathcal{L}_{\mathrm{VGG}}^{(2)}$} \\
& Peak PSNR & Peak SSIM & Peak PSNR & Peak SSIM \\
\hline
$0.1$   & 22.174 & {0.6919} & 22.161 & 0.6869 \\
$1$     & 22.220 & 0.6898 & 22.201 & 0.6931 \\
$10$    & 22.174 & 0.6916 & 22.184 & 0.6863 \\
$100$   & \textbf{22.274} & 0.6895 & {22.221} & \textbf{0.6937} \\
$1000$  & 22.202 & 0.6847 & 21.413 & 0.6444 \\
\hline
SR3 & 22.211 & 0.6910
    & 22.211 & 0.6910 \\
\hline
\end{tabular}
\caption{Sensitivity study of the VGG loss weight on the BrainWeb $8\times$ task. The $\mathcal{L}_{\mathrm{VGG}}^{(1)}$ model with $\omega_{\mathrm{VGG}}=100$ achieves the best peak PSNR, while the $\mathcal{L}_{\mathrm{VGG}}^{(2)}$ model with $\omega_{\mathrm{VGG}}=100$ achieves the best peak SSIM.}
\label{tab:brainweb_8x_vgg_weight_comparison}
\end{table}

The Brain Tumor MRI $4\times$ results in \Cref{tab:tumor_4x_vgg_weight_comparison} further show that the configuration selected on BrainWeb does not transfer directly to another MRI dataset. The squared-$L^2$ model with $\omega_{\mathrm{VGG}}=10$ attains the highest peak PSNR, while the $L^1$ model with $\omega_{\mathrm{VGG}}=0.1$ attains the highest peak SSIM.

\begin{table}[!htbp]
\centering
\begin{tabular}{c|cc|cc}
\hline
\multirow{2}{*}{$\omega_{\mathrm{VGG}}$}
& \multicolumn{2}{c|}{$\mathcal{L}_{\mathrm{VGG}}^{(1)}$}
& \multicolumn{2}{c}{$\mathcal{L}_{\mathrm{VGG}}^{(2)}$} \\
& Peak PSNR & Peak SSIM & Peak PSNR & Peak SSIM \\
\hline
$0.1$   & 29.166 & \textbf{0.8662} & 29.120 & 0.8633 \\
$1$     & 28.945 & 0.8624 & 29.064 & 0.8605 \\
$10$    & 29.090 & 0.8604 & \textbf{29.391} & 0.8645 \\
$100$   & 29.160 & 0.8616 & 27.385 & 0.8204 \\
$1000$  & 27.748 & 0.8334 & 27.414 & 0.8246 \\
\hline
SR3 & 29.217 & 0.8647
    & 29.217 & 0.8647 \\
\hline
\end{tabular}
\caption{Sensitivity study of the VGG loss weight on the Brain Tumor MRI $4\times$ task. The $\mathcal{L}^{(2)}_{\mathrm{VGG}}$ model with $\omega_{\mathrm{VGG}}=10$ achieves the best PSNR, while the $\mathcal{L}_{\mathrm{VGG}}^{(1)}$ model with $\omega_{\mathrm{VGG}}=0.1$ achieves the best SSIM.} \label{tab:tumor_4x_vgg_weight_comparison}
\end{table}

Overall, this sensitivity discussion suggests that at the current stage, both the VGG loss type and the regularization weight $\omega_{\mathrm{VGG}}$ need to be tuned manually for each experimental setting. The preferred configuration depends not only on the dataset and SR scale factor, but also on the optimization objective. One parameter setting may be preferred when faster early-stage convergence is desired, whereas another may yield better final reconstruction quality in terms of PSNR/SSIM. We note, however, that the present study considers only five relatively widely spaced values of $\omega_{\mathrm{VGG}}$, namely ${0.1,1,10,100,1000}$. A finer search around the most promising values may identify configurations with further improvements in convergence or final reconstruction quality.

\paragraph{RGB facial-image experiments}

In this section, we evaluate the effect of adding a VGG perceptual loss to the SR3 framework on face image super-resolution ($4\times$). Models are trained on the FFHQ dataset and evaluated \emph{cross-dataset} on the CelebA-HQ evaluation set. Unlike the grayscale MRI experiments where both the $\mathcal{L}^{(1)}_{\mathrm{VGG}}$ and $\mathcal{L}^{(2)}_{\mathrm{VGG}}$ losses were compared, only the $\mathcal{L}_{\mathrm{VGG}}^{(1)}$ loss is evaluated for CelebA-HQ; we leave the evaluation of the squared-$L^2$ variant on color datasets to future work.

To systematically determine the optimal regularizing weight, we perform a parameter sensitivity study. We structure our search by first sweeping across a coarse range of broad magnitudes, followed by a fine-grained evaluation of tight local increments:
\begin{equation}
\omega_{\mathrm{VGG}} \in \{0.001, 0.01, 0.02, 0.03, 0.04, 0.05, 0.1, 1, 10\}.
\end{equation}

This sweep focuses on finding the general useful range of the parameter space and identifying the configurations that offer the best balance between training speed and final image reconstruction metrics.

\begin{table}[!htbp]
\centering
\begin{tabular}{c|c|c|c|c}
\hline
$\omega_{\mathrm{VGG}}$ & Peak PSNR & Epoch & Peak SSIM & Epoch \\
\hline
0.001 & 27.907 & 542 & 0.8352 & 619 \\
0.01 & 27.842 & 465 & 0.8355 & 465 \\
0.02 & 27.885 & 697 & 0.8309 & 542 \\
0.03 & 28.146 & 619 & 0.8410 & 619 \\
0.04 & \textbf{28.170} & 465 & \textbf{0.8427} & 465 \\
0.05 & 27.506 & 465 & 0.8253 & 465 \\
0.1 & 27.892 & 542 & 0.8347 & 387 \\
1 & 27.925 & 465 & 0.8298 & 465 \\
10 & 27.963 & 774 & 0.8331 & 774 \\
\hline
SR3 (baseline) & 27.747 & 774 & 0.8292 & 774 \\
\hline
\end{tabular}
\caption{Peak PSNR/SSIM for each
$\omega_{\mathrm{VGG}}$ vs.\ the SR3 baseline on CelebA-HQ. Nearly every weight beats the baseline; the exception is $\omega_{\mathrm{VGG}} = 0.05$, which trades final quality for faster convergence (Section~\ref{subsec:color_celeb}).
$\omega_{\mathrm{VGG}} = 0.04$ gives the best scores while saving over 300 epochs.}
\label{tab:combined_metrics}
\end{table}

Table~\ref{tab:combined_metrics} summarizes the peak quantitative metrics and their corresponding training epochs. The results show that perceptual regularization generally improves the final reconstruction performance over the SR3 baseline. Nearly all tested configurations outperform the baseline peak values of 27.747~dB PSNR and 0.8292 SSIM, demonstrating the effectiveness of the proposed perceptual regularization.

Larger weights, such as $\omega_{\mathrm{VGG}} \ge 1$, show diminishing returns relative to the optimal setting, although they still outperform the baseline in both metrics. Looking closely at the fine-grained settings, a critical trade-off emerges between early training acceleration and final peak quality. Specifically, the higher weight configuration of $\omega_{\mathrm{VGG}} = 0.05$ demonstrates very strong early convergence capabilities, allowing the network to quickly climb and plateau at epoch 465. However, this heavy regularization ultimately restricts the model's final capacity, causing the peak scores to drop significantly below the baseline to 27.506~dB.

Among the tested settings, $\omega_{\mathrm{VGG}} = 0.04$ provides the best balance between early convergence and final reconstruction quality. Although its early convergence is more moderate than that of the $\omega_{\mathrm{VGG}} = 0.05$ variant, it achieves the highest peak PSNR of 28.170~dB and peak SSIM of 0.8427 at epoch 465. Compared with SR3, whose peak performance is reached at epoch 774, this setting achieves improved reconstruction quality while reaching its peak performance more than 300 epochs earlier.

In conclusion, integrating the VGG perceptual loss generally improves final image reconstruction quality while accelerating convergence compared to the baseline. Although over-regularization or insufficient weight leads to minimal performance gains or localized degradation, our evaluation successfully establishes the effective operational range of the parameter space. Ultimately, $\omega_{\mathrm{VGG}} = 0.04$ provides the optimal balance, delivering peak quantitative fidelity while maximizing training efficiency.

\section{Conclusion and Future Work}
Image super-resolution remains a challenging inverse problem because image degradations often remove high-frequency information that cannot be recovered directly in many real applications. In this work, we propose a perceptually regularized diffusion framework for single-image super-resolution, built on the SR3 backbone, that augments the standard noise-prediction objective with a VGG-based perceptual regularization term. By penalizing feature-space discrepancy between the reconstructed and ground-truth HR images, the regularization encourages the recovery of perceptually meaningful structures without modifying the diffusion architecture or inference procedure. Numerical experiments on benchmark datasets, including grayscale MRI and color facial images, have shown that the proposed regularization accelerates training convergence and yields more accurate recovery of fine features and textures than the baseline SR3 method. These empirical findings are consistent with the gradient flow interpretation, developed in Section~\ref{sec:gradient}, which shows that the perceptual term introduces anisotropic curvature into the training objective along structurally informative directions. In addition, studies on the parameter selection and stability further confirm the robustness and reproducibility of the proposed framework. Future work includes the development of learnable and multiscale regularization strategies, transfer learning, methods for reducing color shifts, and extensions to other image restoration problems such as denoising and inpainting.

\section*{Acknowledgments}
The authors would like to thank the NSF grant DMS-2520375 that supported the Research Collaboration Workshop in Science of Data and Mathematics (WiSDM) at the University of North Carolina, Chapel Hill, during August 4--8, 2025.

\section*{Author Contributions}
Conceptualization, JQ \& WG; Methodology, JQ \& WG; Software, CW \& PV \& MW; Validation, CW \& PV \& MW; Formal Analysis, MW \& YL; Investigation, CW \& MW; Data Curation, CW \& PV; Writing -- Original Draft Preparation, CW \& YL \& JQ \& YL \& WG; Writing -- Review \& Editing, CW \& PV \& YL \& JQ \& YL \& WG; Supervision - JQ \& YL \& WG; Project Administration - JQ \& WG.

\section*{Competing Interests}
The authors declare no competing interests.

\bibliographystyle{unsrt}
\bibliography{ref}

@inproceedings{ledig2017photo,
  title={Photo-realistic single image super-resolution using a generative adversarial network},
  author={Ledig, Christian and Theis, Lucas and Husz{\'a}r, Ferenc and Caballero, Jose and Cunningham, Andrew and Acosta, Alejandro and Aitken, Andrew and Tejani, Alykhan and Totz, Johannes and Wang, Zehan and others},
  booktitle={Proceedings of the IEEE conference on computer vision and pattern recognition},
  pages={4681--4690},
  year={2017}
}

@inproceedings{wang2018esrgan,
  title={{ESRGAN}: Enhanced super-resolution generative adversarial networks},
  author={Wang, Xintao and Yu, Ke and Wu, Shixiang and Gu, Jinjin and Liu, Yihao and Dong, Chao and Qiao, Yu and Change Loy, Chen},
  booktitle={Proceedings of the European conference on computer vision (ECCV) workshops},
  pages={0--0},
  year={2018}
}

@inproceedings{berrada2025boosting,
  title={Boosting latent diffusion with perceptual objectives},
  author={Berrada, Tariq and Astolfi, Pietro and Hall, Melissa and Havasi, Marton and Benchetrit, Yohann and Romero-Soriano, Adriana and Alahari, Karteek and Drozdzal, Michal and Verbeek, Jakob},
  booktitle={International Conference on Learning Representations},
  year={2025}
}

@article{ho2020denoising,
  title={Denoising diffusion probabilistic models},
  author={Ho, Jonathan and Jain, Ajay and Abbeel, Pieter},
  journal={Advances in neural information processing systems},
  volume={33},
  pages={6840--6851},
  year={2020}
}

@article{Simonyan2015VGG,
 title = {Very deep convolutional networks for large-scale image recognition},
 type = {article},
 year = {2015},
 pages = {1-14},
 id = {66023775-a7b1-3c01-9142-bdf478ea9c1d},
 created = {2021-11-01T10:14:38.920Z},
 file_attached = {true},
 profile_id = {235249c2-3ed4-314a-b309-b1ea0330f5d9},
 group_id = {1ff583c0-be37-34fa-9c04-73c69437d354},
 last_modified = {2022-07-21T08:32:29.142Z},
 starred = {false},
 authored = {false},
 confirmed = {true},
 hidden = {false},
 citation_key = {Simonyan2015},
 folder_uuids = {1853f94b-7af1-40fa-b068-4758e9a02bc4},
 private_publication = {false},
 bibtype = {article},
 author = {Simonyan, Karen and Zisserman, Andrew},
 journal = {3rd International Conference on Learning Representations, ICLR 2015 - Conference Track Proceedings}
}

@article{saharia2022image,
  title={Image super-resolution via iterative refinement},
  author={Saharia, Chitwan and Ho, Jonathan and Chan, William and Salimans, Tim and Fleet, David J and Norouzi, Mohammad},
  journal={IEEE transactions on pattern analysis and machine intelligence},
  volume={45},
  number={4},
  pages={4713--4726},
  year={2022},
  publisher={IEEE}
}

@article{li2022srdiff,
  title={{SRDiff}: Single image super-resolution with diffusion probabilistic models},
  author={Li, Haoying and Yang, Yifan and Chang, Meng and Chen, Shiqi and Feng, Huajun and Xu, Zhihai and Li, Qi and Chen, Yueting},
  journal={Neurocomputing},
  volume={479},
  pages={47--59},
  year={2022},
  publisher={Elsevier}
}

@article{moser2024diffusion,
  title={Diffusion models, image super-resolution, and everything: A survey},
  author={Moser, Brian B and Shanbhag, Arundhati S and Raue, Federico and Frolov, Stanislav and Palacio, Sebastian and Dengel, Andreas},
  journal={IEEE Transactions on Neural Networks and Learning Systems},
  year={2024},
  publisher={IEEE}
}

@article{yang2023diffusion,
  title={Diffusion models: A comprehensive survey of methods and applications},
  author={Yang, Ling and Zhang, Zhilong and Song, Yang and Hong, Shenda and Xu, Runsheng and Zhao, Yue and Zhang, Wentao and Cui, Bin and Yang, Ming-Hsuan},
  journal={ACM computing surveys},
  volume={56},
  number={4},
  pages={1--39},
  year={2023},
  publisher={ACM New York, NY, USA}
}

@article{yue2023resshift,
  title={ResShift: Efficient diffusion model for image super-resolution by residual shifting},
  author={Yue, Zongsheng and Wang, Jianyi and Loy, Chen Change},
  journal={Advances in Neural Information Processing Systems},
  volume={36},
  pages={13294--13307},
  year={2023}
}

@inproceedings{shang2024resdiff,
  title={Resdiff: Combining cnn and diffusion model for image super-resolution},
  author={Shang, Shuyao and Shan, Zhengyang and Liu, Guangxing and Wang, LunQian and Wang, XingHua and Zhang, Zekai and Zhang, Jinglin},
  booktitle={Proceedings of the AAAI Conference on Artificial Intelligence},
  volume={38},
  number={8},
  pages={8975--8983},
  year={2024}
}

@inproceedings{khaledyan2020low,
  title={Low-cost implementation of bilinear and bicubic image interpolation for real-time image super-resolution},
  author={Khaledyan, Donya and Amirany, Abdolah and Jafari, Kian and Moaiyeri, Mohammad Hossein and Khuzani, Abolfazl Zargari and Mashhadi, Najmeh},
  booktitle={2020 IEEE Global Humanitarian Technology Conference (GHTC)},
  pages={1--5},
  year={2020},
  organization={IEEE}
}

@inproceedings{liu2013directional,
  title={Directional bicubic interpolation—A new method of image super-resolution},
  author={Liu, Jing and Gan, Zongliang and Zhu, Xiuchang},
  booktitle={3rd International Conference on Multimedia Technology (ICMT-13)},
  pages={463--470},
  year={2013},
  organization={Atlantis Press}
}

@inproceedings{kingma2014vae,
  title     = {Auto-Encoding Variational {Bayes}},
  author    = {Kingma, Diederik P and Welling, Max},
  booktitle = {International Conference on Learning Representations},
  year      = {2014}
}

@inproceedings{goodfellow2014gan,
  title     = {Generative Adversarial Nets},
  author    = {Goodfellow, Ian and Pouget-Abadie, Jean and Mirza, Mehdi and Xu, Bing and Warde-Farley, David and Ozair, Sherjil and Courville, Aaron and Bengio, Yoshua},
  booktitle = {Advances in Neural Information Processing Systems},
  year      = {2014}
}

@inproceedings{rezende2015flows,
  title     = {Variational Inference with Normalizing Flows},
  author    = {Rezende, Danilo and Mohamed, Shakir},
  booktitle = {International Conference on Machine Learning},
  year      = {2015}
}

@inproceedings{sohl2015deep,
  title     = {Deep Unsupervised Learning using Nonequilibrium Thermodynamics},
  author    = {Sohl-Dickstein, Jascha and Weiss, Eric and Maheswaranathan, Niru and Ganguli, Surya},
  booktitle = {International Conference on Machine Learning},
  year      = {2015}
}

@inproceedings{song2021scorebased,
  title     = {Score-Based Generative Modeling through Stochastic Differential Equations},
  author    = {Song, Yang and Sohl-Dickstein, Jascha and Kingma, Diederik P. and Kumar, Abhishek and Ermon, Stefano and Poole, Ben},
  booktitle = {International Conference on Learning Representations},
  year      = {2021}
}

@article{Cocosco1997BrainWebOI,
  title={{BrainWeb}: Online Interface to a {3D MRI} Simulated Brain Database},
  author={Chris A. Cocosco and Vasken Kollokian and Remi K.-S. Kwan and Alan C. Evans},
  journal={NeuroImage},
  year={1997},
  url={https://api.semanticscholar.org/CorpusID:14864257}
}

@article{wiatrak2019stabilizing,
  title={Stabilizing generative adversarial networks: A survey},
  author={Wiatrak, Maciej and Albrecht, Stefano V and Nystrom, Andrew},
  journal={arXiv preprint arXiv:1910.00927},
  year={2019}
}

@inproceedings{xie2023desra,
  title={{DeSRA}: Detect and Delete the Artifacts of {GAN}-based Real-World Super-Resolution Models},
  author={Xie, Liangbin and Wang, Xintao and Chen, Xiangyu and Li, Gen and Shan, Ying and Zhou, Jiantao and Dong, Chao},
  booktitle={International Conference on Machine Learning},
  pages={38204--38226},
  year={2023},
  organization={PMLR}
}

@inproceedings{johnson2016perceptual,
  title={Perceptual losses for real-time style transfer and super-resolution},
  author={Johnson, Justin and Alahi, Alexandre and Fei-Fei, Li},
  booktitle={European conference on computer vision},
  pages={694--711},
  year={2016},
  organization={Springer}
}

@ARTICLE{Kwan1999MRI,
  author={Kwan, R.K.-S. and Evans, A.C. and Pike, G.B.},
  journal={IEEE Transactions on Medical Imaging}, 
  title={{MRI} simulation-based evaluation of image-processing and classification methods}, 
  year={1999},
  volume={18},
  number={11},
  pages={1085-1097},
  doi={10.1109/42.816072}}

@inproceedings{kwan1996extensible,
  title={An extensible {MRI} simulator for post-processing evaluation},
  author={Kwan, Remi K-S and Evans, Alan C and Pike, G Bruce},
  booktitle={International conference on visualization in biomedical computing},
  pages={135--140},
  year={1996},
  organization={Springer}
}

@article{Collins1998Design,
  title={Design and construction of a realistic digital brain phantom},
  author={D. Louis Collins and Alex P. Zijdenbos and Vasken Kollokian and John G. Sled and Noor Jehan Kabani and Colin J. Holmes and Alan C. Evans},
  journal={IEEE Transactions on Medical Imaging},
  year={1998},
  volume={17},
  pages={463-468},
  url={https://api.semanticscholar.org/CorpusID:30712309}
}

@inproceedings{karras2019stylebased,
  title     = {A Style-Based Generator Architecture for Generative Adversarial Networks},
  author    = {Karras, Tero and Laine, Samuli and Aila, Timo},
  booktitle = {Proceedings of the IEEE/CVF Conference on Computer Vision and Pattern Recognition},
  pages     = {4401--4410},
  year      = {2019}
}

@inproceedings{liu2015faceattributes,
 title = {Deep Learning Face Attributes in the Wild},
 author = {Liu, Ziwei and Luo, Ping and Wang, Xiaogang and Tang, Xiaoou},
 booktitle = {Proceedings of International Conference on Computer Vision (ICCV)},
 year = {2015} 
}

@inproceedings{karras2017progressive,
  title={Progressive growing of {GANs} for improved quality, stability, and variation},
  author={Karras, Tero and Aila, Timo and Laine, Samuli and Lehtinen, Jaakko},
  journal={International Conference on Learning Representations (ICLR)},
  year={2018}
}

@article{wang2020deep,
  title={Deep learning for image super-resolution: A survey},
  author={Wang, Zhihao and Chen, Jian and Hoi, Steven CH},
  journal={IEEE transactions on pattern analysis and machine intelligence},
  volume={43},
  number={10},
  pages={3365--3387},
  year={2020},
  publisher={IEEE}
}

@article{chen2022real,
  title={Real-world single image super-resolution: A brief review},
  author={Chen, Honggang and He, Xiaohai and Qing, Linbo and Wu, Yuanyuan and Ren, Chao and Sheriff, Ray E and Zhu, Ce},
  journal={Information Fusion},
  volume={79},
  pages={124--145},
  year={2022},
  publisher={Elsevier}
}

@article{rudin1992nonlinear,
  title={Nonlinear total variation based noise removal algorithms},
  author={Rudin, Leonid I and Osher, Stanley and Fatemi, Emad},
  journal={Physica D: nonlinear phenomena},
  volume={60},
  number={1-4},
  pages={259--268},
  year={1992},
  publisher={Elsevier}
}

\end{document}